\UseRawInputEncoding
\documentclass[namedreferences]{solarphysics}
\usepackage[hyperref, optionalrh, solaromanenum]{spr-sola-addons} 
\usepackage{graphicx} 
\usepackage{natbib}                  
\usepackage{amssymb}                    
\usepackage{color}                       
\usepackage{breakurl}
\usepackage{array}                   

\begin{document}


\begin{opening}

\title{Extracting Hale Cycle Related Components from Cosmic-Ray Data Using Principal Component Analysis}

%
\author[addressref={aff1},corref,email={jojuta@gmail.com}]{\inits{J.J.}\fnm{Jouni}~\lnm{Takalo}}

\institute{$^{1}$ Space Physics and Astronomy Research Unit, University of Oulu, POB 3000, FIN-90014, Oulu, Finland}
%
\runningauthor{J.J. Takalo}
\runningtitle{Extracting Hale cycle from CR using PCA}



\begin{abstract}

We decompose the monthly cosmic-ray data, using several neutron monitor count rates, of Cycles 19\,--\,24 with principal component analysis (PCA). We show using different cycle limits that the first and second PC of cosmic-ray (CR) data explain 77\,--\,79\% and 13\,--\,15\% of the total variation of the Oulu CR Cycles 20\,--\,24(C20\,--\,C24), 73\,--\,77\% and 13\,--\,17\% of the variation of Hermanus C20\,--\,C24, and 74\,--\,78\% and 17\,--\,21\% of the Climax C19\,--\,C22, respectively. The PC1 time series of the CR Cycles 19\,--\,24 has only one peak in its power spectrum at the period 10.95 years, which is the average solar cycle period for the interval SC19\,--\,SC24. The PC2 time series of the same cycles has a clear peak at period 21.90 (Hale cycle) and another peak at 1/3 of that period with no peak at the solar cycle period.
We show that the PC2 of the CR is essential in explaining the differences in the intensities of the even and odd cycles of the CR. The odd cycles have positive phase in the first half and negative phase in the second half of their PC2. This leads to slow decrease of the intensity in the beginning of the cycle and flat minimum for the odd cycles. On the contrary, for the even cycles the phases are vice versa and this leads to faster decrease and more rapid recovery in the CR intensity of the cycle. As a consequence the even cycles have more peak-like structure.
The only exceptions of this rule are Cycles 23 and 24 such the former has almost zero line PC2, and the latter has similar PC2 than the earlier odd cycles. The reason for this may be that the aforementioned rule is only valid for grand solar maximum cycles or that there is a phase shift going on in the CR overall shape.  
These results are confirmed with skewness-kurtosis (S\,--\,K) analysis. Furthermore, S\,--\,K shows that other even and odd cycles, except Cycle 21, are on the regression line with correlation coefficient 0.85. The Cycles 21 of all calculated eight stations are compactly located in the S\,--\,K coordinate system and have smaller skewnesses and higher kurtoses than the odd Cycles 23.

\end{abstract}


\end{opening}

\section{Introduction}

The neutron monitor count rates have known period of about 11-years, i.e. the average length of solar cycle. However, it has been noticed that the cosmic-ray data experiences also 22-year or so-called Hale cycle, which is the magnetic polarity cycle of the sun \citep{Webber_1988, Mavromichalaki_1997, VanAllen_2000, Thomas_2014, Kane_2014, Ross_2019}.

\cite{Webber_1988} found a sharply peaked CR intensity maximum in 1987, similar to that observed 22 years earlier in 1965, in contrast to the flatter maximum observed between 1972 and 1977 and earlier in 1952-1954. In addition, they found that the neutron monitor (NM) count rate is about 1.5\% higher at the times of maximum in 1965 and 1987 as compared with the 1972-1977 period.

\cite{Mavromichalaki_1997} reported that for the CR odd cycles (C19 and C21) the decreasing of the intensity is slow and peaks shortly close to the cycle minimum and has saddle-like shape. In addition odd cycles have a long-recovery time and long time lag between CR and solar sunspot number (SSN). On the other hand, even CR cycles (C20 and C22) have a two minima structure and overall peak-type structure. Furthermore, the even cycles have rapid recovery phase and short time lag to the SSN. They also studied the dependence of the CR as a function of SSN and found that the hysteresis curve is more elliptic for the odd cycles than for the even cycles.

\cite{VanAllen_2000} confirms that there is a striking difference between such modulation loops. i.e. hysteresis curves between CR and SSN for solar activity Cycles 19 and 21 and those for Cycles 20 and 22. The loops for Cycles 19 and 21 are broad ovals whereas those for Cycles 20 and 22 are nearly flat.
Furthermore, the cosmic-ray intensity decreases more rapidly as the SSN increases following solar activity minima when the solar polar magnetic parameter A is negative (magnetic moment and solar rotation axis are anti-parallel) than when A is positive (magnetic moment and rotation axis are parallel). He argues that these facts give some support to a significant role of gradient and curvature drifts in CR transport in heliosphere.

\cite{Thomas_2014} state that the onset of the peak cosmic-ray flux at Earth occurs earlier during A$>$0 (odd) cycles than for A$<$0 (even) cycles and hence the peak being more dome-like for A$>$0 and more sharply peaked for A$<$0. They also demonstrate that these polarity-dependent heliospheric differences are evident during the space-age but much less clear in earlier data: using geomagnetic reconstructions, they show that for the period of 1905\,--\,1965, alternate polarities do not give as significant difference during the declining phase of the solar cycle. They suggest that the 22-year cycle in cosmic-ray flux is at least partly the result of direct modulation by the heliospheric magnetic field and that this effect may be primarily limited to the grand solar maximum of the space-age.

\cite{Owens_2015} found that approximately 22-year Hale cycle means that odd- and even-numbered Schwabe, i.e. solar/sunspot cycles, are associated with different patterns in the galactic cosmic-ray intensity. They used recent $^{10}$Be concentration measurements in their analysis.

\cite{Thomas_2017} observe a difference in the phase and amplitude of the daily variation in neutron monitor count rates with the Sun’s magnetic polarity. The timing varies in a 22-year Hale cycle with maxima and minima latest in the day at solar minimum in A$<$0 cycles and earliest at solar minimum during A$>$0 cycles.

\cite{Kane_2014} found that hysteresis plots between cosmic-ray (CR) intensity (recorded at the Climax station) and sunspot number show broad loops in odd cycles (19, 21, and 23) and narrow loops in even cycles (20 and 22 ). However, in the even cycles, the loops are not narrow throughout the whole cycle; around the sunspot-maximum period, a broad loop is seen. Only in the rising and declining phases, the loops are narrow in even cycles. Thus, the differences between odd and even cycles are not significant throughout the whole cycle. In the recent even Cycle 24, hysteresis plots show a preliminary broadening near the first sunspot maximum of 2012. 

\cite{Ross_2019} show also that plots of SSN versus GCR reveal a clear difference between the odd and even-numbered cycles. Linear and elliptical models have been fit to the data, with the linear fit and elliptical model proving the more suitable model for even and odd solar-activity cycles, respectively, in agreement with the earlier studies. They study especially the lag and hysteresis of CR Cycle 24. They found that CR Cycle 24 has a slightly longer lag from SSN than previous even cycles. They suggest that the extended lag in Cycle 24 compared to previous even-numbered cycles is likely due to the deep, extended minimum between Cycles 23 and 24, and the low maximum activity of Cycle 24
 
\cite{Mishra_2008} found that the IMF strength (vB) shows a weak negative correlation (-0.35) with cosmic-rays for solar cycle 20, 
and a good anti-correlation for solar cycles 21\,–\,23 (-0.76, -0.69) with the cosmic-ray intensity.

\cite{Hempelmann_2012} used Fourier analysis of the time series and suspected that the cosmic-rays have a link with solar activity following their observations that showed significant peaks of 10.7, 22.4, and 14.9 years.

\cite{Broomhall_2017} modeled the hysteresis plots using both a simple linear model and an ellipse model, because of the difference in the shape of the hysteresis plots for odd and even cycles. Their results of this study tend to support that Cycle 24 follows the same trend as preceding even cycles and is better represented by a straight line rather than an ellipse.

\cite{Oloketuyi_2020} found that CR intensity undergoes 11-year solar cycle within the heliosphere, which is greatly influenced mainly by solar activities. The cycle formed has its peak at the solar minimum and vice-versa. They study also confirmed that the daily sunspot numbers and CRI are negatively correlated (correlation coefficient -0.72 for cycle 23 and -0.73 for cycle 24), i.e anti-correlations observed from the cycles are highly significant. Solar wind speed was, however, found to be uncorrelated with SSN.

In this article we use different methods from the earlier studies of CR cycles. We separate the solar cycle related and Hale cycle related components from the CR flux data using principal component analysis (PCA).  We show that the odd and even cycles differ in their second principal component (PC2). Furthermore, we show that also the skewness and kurtosis of the cycles separate even cycles, except even Cycle 24, from the odd cycles. This paper is organized as follows. Section 2 presents the data and methods used in this study. In Section 3 we present the results of PC and skewness-kurtosis analyses for CR data of the Cycles 19\,--\,24 and discuss the results.  We also make PC analyses to solar/solar wind data, and compare the results to those of CR data. We give our conclusion in Section 4.

\begin{figure}
\centering
	\includegraphics[width=1.0\textwidth]{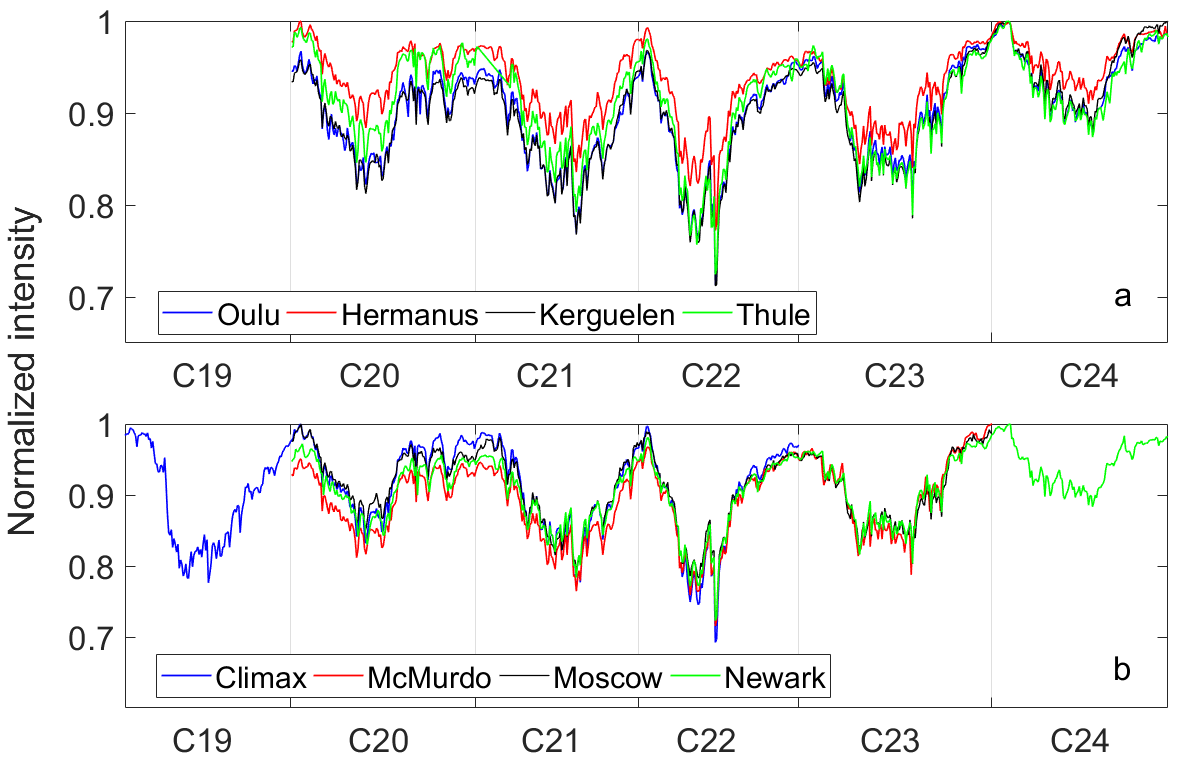}
		\caption{The cosmic-ray data as normalized intensities for eight neutron monitor (NM) stations.}
		\label{fig:Oulu_Climax_ja_muut}
\end{figure}

\section{Data and Methods}

\subsection{Cosmic-Ray Data}

Although cosmic-ray intensity lags somewhat behind the solar sunspot numbers, we have used the dates mentioned in Table 1 for the cosmic-ray data cycles in the first PCA analysis. We used this choice first because the lags from solar cycles vary between different studies \citep{Mavromichalaki_1997, Gupta_2005, Rybansky_2009, Kane_2014, Porta_2018, Iskra_2019, Ross_2019, Koldobskiy_2022}. We also varied the starting dates in order to see if this affects to the results of the PC analysis. The other dates, which we use in this paper are the local maxima between different cosmic-ray cycles. These are shown in Table 2. We used in our analyses only cosmic-ray data, which have at least four total cycles, excepting with a few values missing due to gaps in the measurement. For this reason we chose Oulu CR flux as a primary data, because it consists five whole Cycles 20\,--\,24. We also study Hermanus and Climax (Colorado) CR as another primary data, because the former station is located in the southern hemisphere and the latter is the only reliable station with continuous values for the whole Cycles 19\,--\,22. Note, however, that the early neutron monitors were more simple, called 'IGY' type monitors, and only since 1964 the improved NM64 monitors have been used \citep{Vaisanen_2021}. In comparison for these data we use data from Kerguelen (20\,--\,24), McMurdo (20\,--\,23), Moscow (20\,--\,23), Newark (20\,--\,24) and Thule (20\,--\,24). Figure \ref{fig:Oulu_Climax_ja_muut}a and b show the cosmic-ray intensities at Oulu and Climax neutron monitor (NM) stations for the Cycles 20\,--\,24 and 19\,--\,22, respectively. We have added the other station to these figures also (see captions of the Fig. \ref{fig:Oulu_Climax_ja_muut}). Because the cutoff rigidities and amount of NMs differ in different stations, i.e., the levels of the counts vary a lot \citep{Usoskin_2017, Vaisanen_2021}, we normalize here the CR intensities such that the maximum of each station is one. The normalizing method does not affect the PCA, because we do analyses separately for each station data. Note also from the Fig. \ref{fig:Oulu_Climax_ja_muut} that the maxima of the CR intensities are lagging from the minima of the solar cycles shown as vertical lines. Another normalization, which we have to do, is resampling (interpolation) of the cycles. This is because each cycle of the individual station must have same length to be able to present the cycles as a correlation matrix needed for the PCA. This procedure ensures that each cycle has equal weight to the common cycle shape, although their amplitudes are different. The cycles are, however, reverted to their original lengths and amplitudes after the PCA process.

\begin{table}

\caption{Sunspot cycle lengths and dates of starting minima used in the first analysis.}

\begin{tabular}{ |c|c|c| }
\hline
   Sunspot cycle    &Year and month  &Cycle length  \\
	    number        &of starting min   &(years)     \\ 
\hline

19       &1954 April  & 10.5  \\
20       &1964 October  & 11.7  \\
21       &1976 June & 10.2  \\
22       &1986 September  & 10.1 \\
23       &1996 October  & 12.2  \\
24       &2008 December & 11.0\\
25       &2019 December &  \\ 

\hline

\end{tabular}

\end{table}

\subsection{Principal Component analysis Method}

Principal component analysis is a useful tool in many fields of science including chemometrics \citep{Bro_2014}, data compression \citep{Kumar_2008} and information extraction \citep{Hannachi_2007}. PCA finds combinations of variables, that describe major trends in the data. PCA has earlier been applied, e.g., to studies of the geomagnetic field \citep{Bhattacharyya_2015}, geomagnetic activity \citep{Holappa_2014_2, Takalo_2021b}, ionosphere \citep{Lin_2012}, the solar background magnetic field \citep{Zharkova_2015}, variability of the daily cosmic-ray count rates \citep{Okpala_2014}, and atmospheric correction to cosmic-ray detectors \citep{Savic_2019}. As far as we know, this is the first time to study and compare cosmic-ray cycles using PCA.

To this end, we estimate that the average length of the cycle is 133 months, and use it as a representative cosmic-ray cycle. We first resample the monthly cosmic-ray counts such that all cycles have the same length of 133 time steps (months), i.e about the average length of the Solar Cycles 20-24 \citep{Takalo_2018, Takalo_2020a, Takalo_2021b}. This effectively elongates or abridges the cycles to the same length. Before applying the PCA method to the resampled cycles we standardize each individual cycle to have zero mean and unit standard deviation. This guarantees that all cycles will have the same weight in the study of their common shape. Then after applying the PCA method to these resampled and standardized cycles, we revert the cycle lengths and amplitudes to their original values.

As said earlier, we standardize each individual cycle of CR flux to have zero mean and unit standard deviation. Standardized data are then collected into the columns of the matrix $X$, which can be decomposed as \citep{Hannachi_2007, Holappa_2014_1, Takalo_2018}

\begin{equation}
	X = U\:D\;V^{T}  \     ,
\end{equation}

where $U$ and $V$ are orthogonal matrices, $V^{T}$ a transpose of matrix $V$, and $D$ a diagonal matrix 
	$D= diag\left(\lambda_{1},\lambda_{2},...,\lambda_{n}\right)$
with $\lambda_{i}$ the $i^{th}$ singular value of matrix $X$. The principal component are obtained as the the column vectors of

\begin{equation}
P  = U\!D.
\end{equation}
	
The column vectors of the matrix $V$ are called empirical orthogonal functions (EOF) and they represent the weights of each principal component in the decomposition of the original normalized data of each cycle $X_{i}$, which can be approximated as

\begin{equation}
	X_{i} = \sum^{N}_{j=1} \:P_{ij}\:V_{ij} \   ,
\end{equation}

where j denotes the $j^{th}$ principal component (PC). The explained variance of each PC is proportional to square of the corresponding singular value
$\lambda_{i}$. Hence the $i^{th}$
PC explains a percentage
\begin{equation}
\frac{\lambda^{2}_{i}}{\sum^{n}_{k=1}\!\lambda^{2}_{k}} \cdot\:100\%
\end{equation}
of the variance in the data.

\subsection{Skewness and Kurtosis}

The moments are a set of constants that represent some important properties of the distributions. The most common are first moment and second moments, i.e. expectation value (mean value) and variance. Two other important measures are the coefficients of skewness and kurtosis, i.e. third and fourth moments. The skewness measures the degree of asymmetry of the distribution, whereas kurtosis measures the degree of flatness of the distribution. The skewness can be defined as
\begin{equation}
\beta_{1}\,=\,\frac{\sum^{N}_{i=1}\left(Y_{i}-\overline{Y}\right)^{3}/N}{s^{3}}  ,
\end{equation}

where $\overline{Y}$ is the mean, s is the standard deviation, i.e. square root of variance, and N is the number of data points. If skewness is positive/negative, the distribution is skewed to the right/left.

The kurtosis is defined as
\begin{equation}
\beta_{2}\,=\,\frac{\sum^{N}_{i=1}{\left(Y_{i}-\overline{Y}\right)^{4}}/N}{s^{4}}  .
\end{equation}

Kurtosis is sometimes thought to tell about the peakedness of the distribution, but actually it tells about the tails and flatness of the distribution. If kurtosis has high/low value it has heavy/light tails \citep{Krishnamoorthy_2006}. The skewness-kurtosis pair of normal distribution is (0,3). That is why sometimes the coefficient 'excess kurtosis' is used, i.e. the aforementioned kurtosis-3.

\begin{table}

\caption{Cosmic-ray cycle lengths and dates of starting minima used in the second and third analyses.}
\begin{tabular}{ |c|c|c|c|c| }

\hline
   Cycle         &Year and month      &Cycle length      &Year and month      &Cycle length \\
	 number   &of 2$^{nd}$ starting min    &(years)    &of 3$^{rd}$ starting min   &(years)   \\
\hline 

19       &1954 August  & 10.8    &1955 April  & 10.8  \\
20       &1965 May  & 11.4       &1965 December & 11.4 \\   
21       &1976 August & 10.5     &1977 April & 10.0 \\
22       &1987 February  & 10.0  &1987 April & 10.8\\
23       &1997 January  & 12.2   &1998 January & 11.8\\
24       &2009 March & 10.4      &2009 December & 10.3\\
25       &2019 September  &      &2020 March\\

\hline

\end{tabular}

\end{table}

\section{Results and Discussion}

\subsection{PC Analyses}

\begin{figure}
	\centering
	\includegraphics[width=0.9\textwidth]{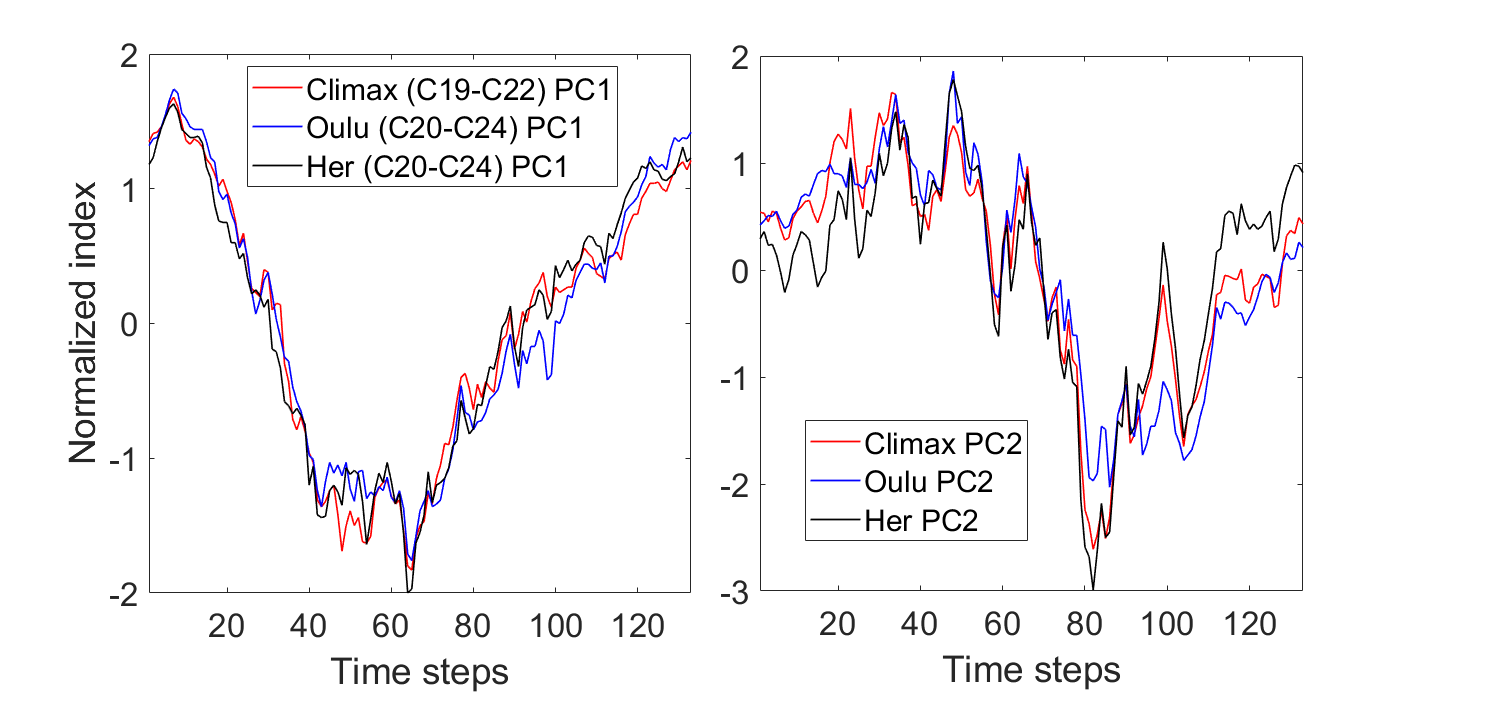}
		\caption{The PC1s in a) and PC2s in b) for the Oulu, Hermanus and Climax cosmic-ray intensities of the first PC analysis.}
		\label{fig:PC1_PC2s}
\end{figure}

Using Oulu, Hermanus and Climax cosmic-ray data, a PCA analysis was conducted by equalizing the cycles to 133 time steps (months) to get the two main principal components shown in Fig. \ref{fig:PC1_PC2s}. The first and second PC explain 79.0\,\% and 13.3\,\% of the total variation of the Oulu C20-C24 data, 77.0\% and 13.2\% of the variation of Hermanus C20-C24  and 76.7\,\% and 18.8\,\% of the Climax C19-C22 data. Hence the two main PCs account together for 92.3\,\%, 90.2\% and 95.5\,\% of the total variation of CR data for Oulu, Hermanus and Climax, respectively. Note, especially, that the PC2 explains almost one fifth of the Climax CR intensity.

\begin{figure}
	\centering
	\includegraphics[width=0.9\textwidth]{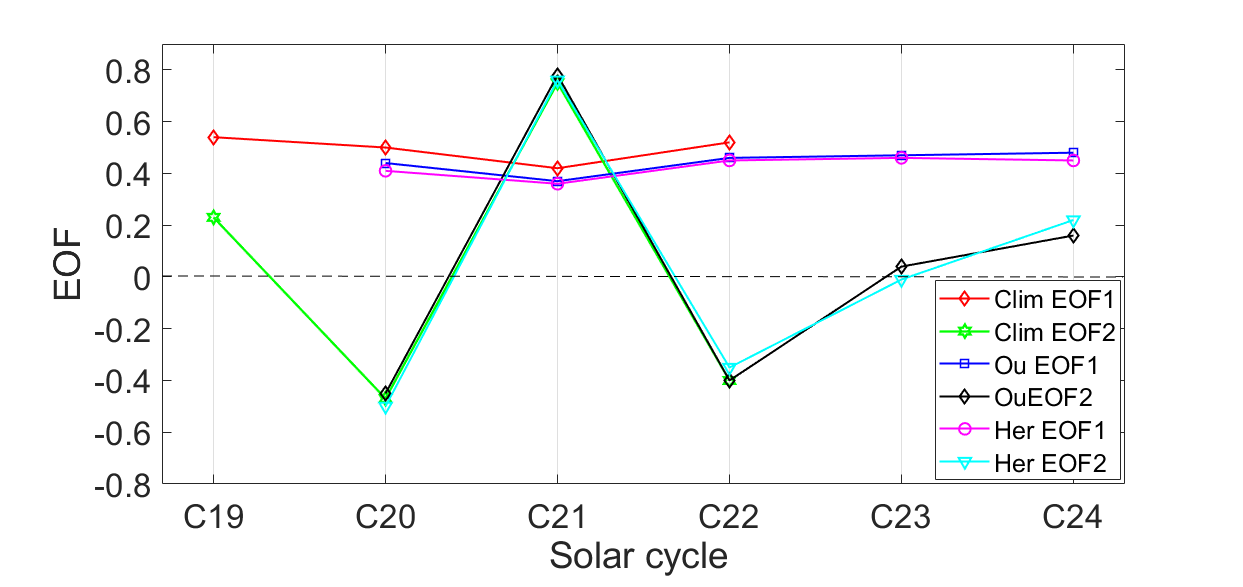}
		\caption{The EOF1s in a) and EOF2s in b) of the Oulu, Hermanus and Climax CR PC analyses.}
		\label{fig:EOF1_EOF2}
\end{figure}

Figure \ref{fig:EOF1_EOF2} shows the corresponding EOFs of the Oulu, Hermanus and Climax CR intensities. Note the almost sawtooth shape of the EOF2. The EOF2s of odd cycles are positive and EOFs of even cycles are negative, except for EOF2 of Cycle 24 of Oulu and Hermanus which have opposite sign to those for earlier even cycles. Note also that EOF2 for Hermanus Cycle 23 is slightly negative in contrast to other odd cycles. Looking the shape of PC2s of Fig. \ref{fig:PC1_PC2s}, we note that a positive EOF2 means positive phase in the first half of the cycle, and a negative phase in the second half of the cycle. For negative EOF2 the phases are opposite. Note the similarity of the EOF1 and EOF2 in the cycles, which are common for all stations. EOF1s for Climax are little higher, because the PCA is calculated from only four cycles, while Oulu and Hermanus PCA are calculated from five cycles. It is essential, that the cycle 21 has lowest weight to all PC1s, but highest weight to all PC2s. (Later I call PC2 phase positive if it starts with positive phase (as in Fig. \ref{fig:PC1_PC2s}b), and negative if it starts with negative phase.)

\begin{figure}
	\centering
	\includegraphics[width=1.0\textwidth]{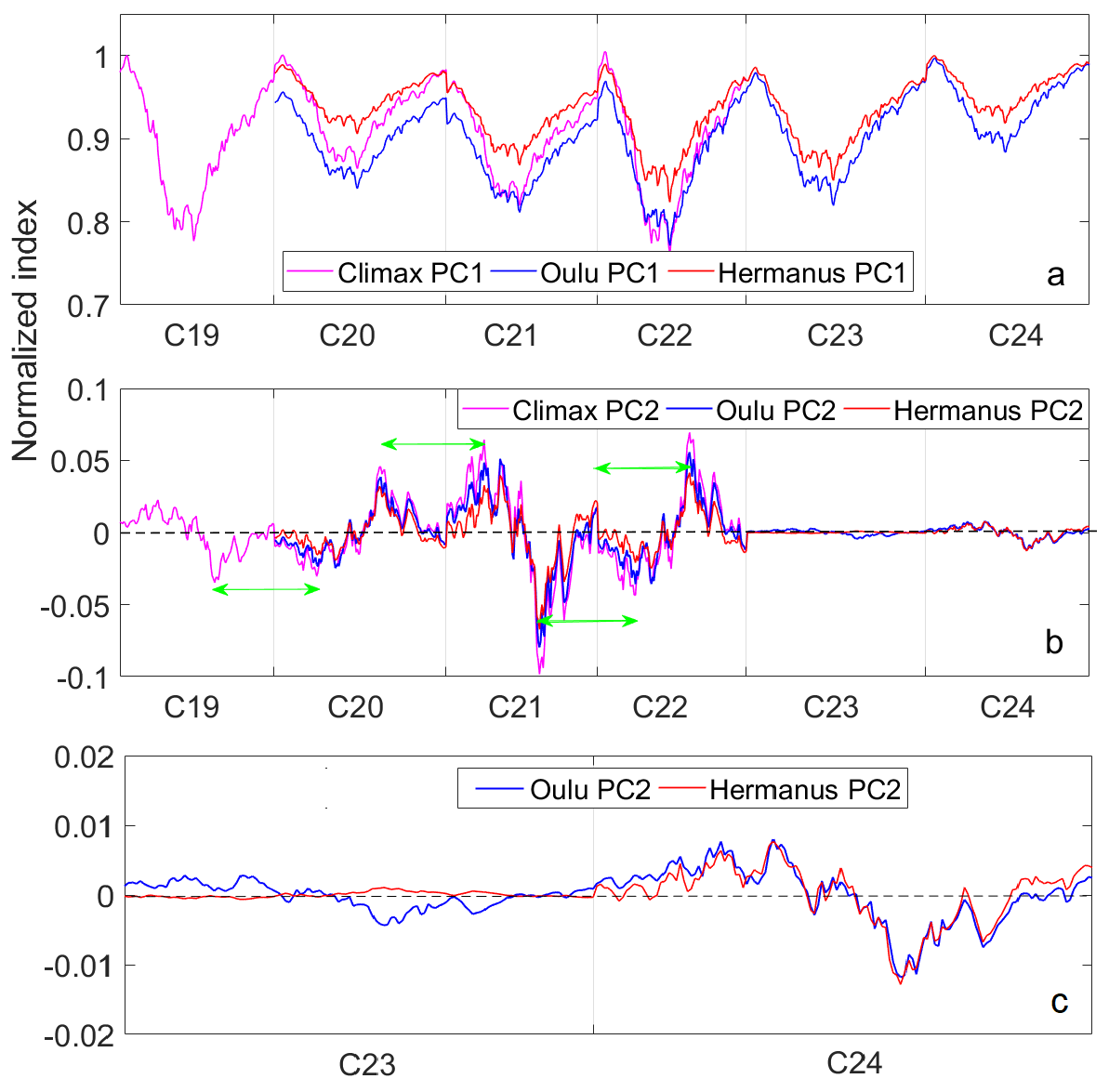}
		\caption{a PC1 and b) PC2 time series of the PCA of CR intensities of Oulu, Hermanus and Climax for Cycles 19\,--\,24. c) magnification of C23 and C24 for Oulu and Hermanus.}
		\label{fig:PC_timeseries_first}
\end{figure}

\begin{figure}
	\centering
	\includegraphics[width=1.0\textwidth]{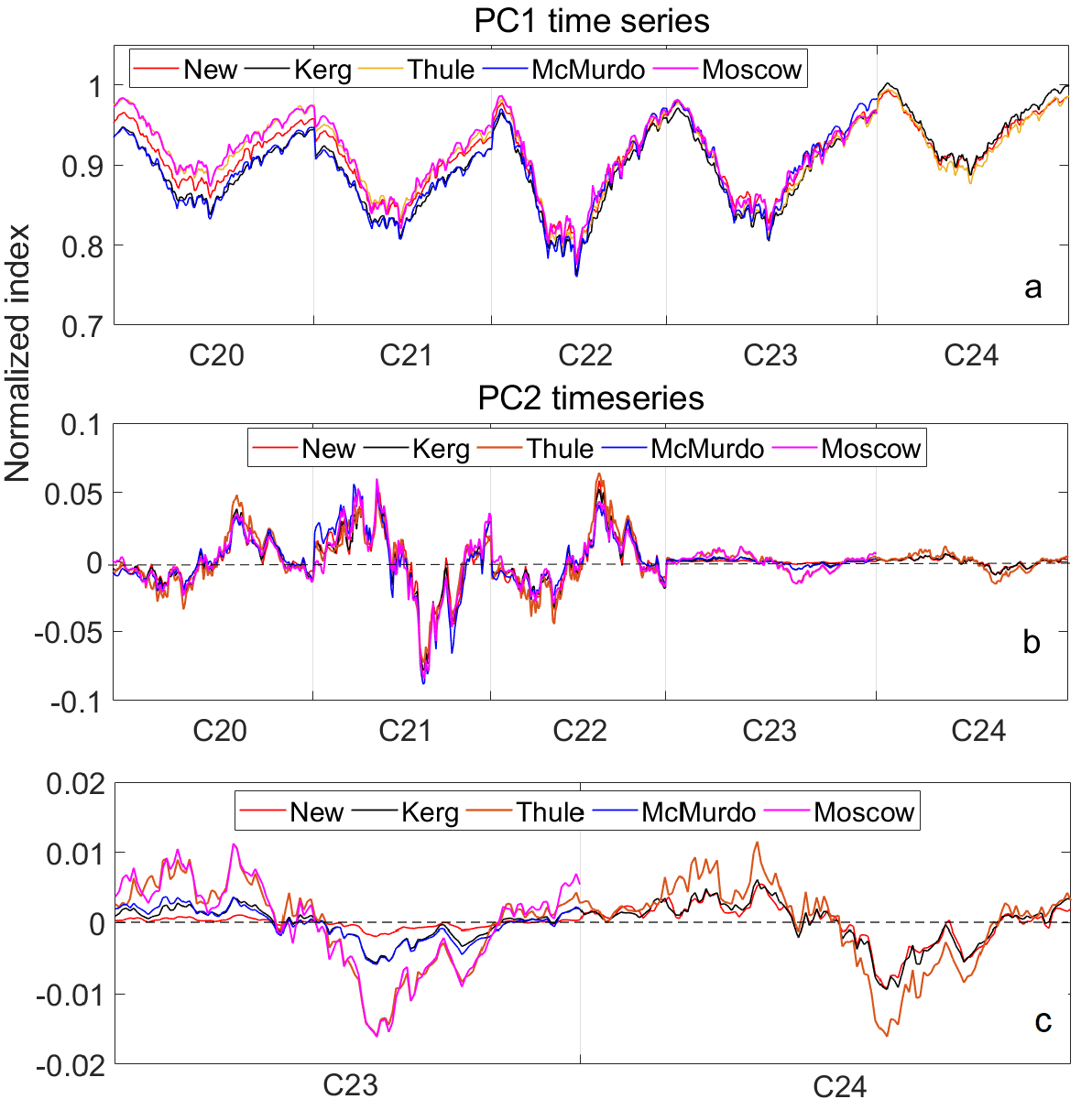}
		\caption{a PC1 and b) PC2 time series of the PCA of CR intensities of Newark, Kerguelen, Thule, McMurdo and Moscow.  c) magnification of cycles 23 and 24.}
		\label{fig:PC_timeseries_rest}
\end{figure}

Reverting the PC1, PC2 cycles of Oulu, Hermanus and Climax CR back to their original length and concatenating them we get first and second principal component time series for the CR cycles shown in Fig. \ref{fig:PC_timeseries_first}a and b, respectively (we still use the same normalization as in the Fig. \ref{fig:Oulu_Climax_ja_muut}). It is evident, that the PC1 is dominated by the sunspot cycle related period. Interestingly, the largest size of PC1 is almost equal for Cycles 19 and 22, and smallest for Cycles 20 and 24. 

In contrast with PC1, the PC2 shows clear Hale cycle period (see Fig. \ref{fig:PC_timeseries_first}b). Note that the consecutive even-odd cycle pairs show similar structure for two Hale cycles. Hale cycle is traditionally referred as even-odd cycle pair \citep{Gnevyshev_1948,Wilson_1988, Makarov_1994, Cliver_1996, Takalo_2021b}. However, we start here with odd Cycle 19, because this is the first cycle with cosmic-ray measurements. Note that cycle pairs 19\,--\,20 and 21\,--\,22 have opposite phases, i.e., first cycle is positive and second cycle negative. Note also a very strong peak-to-peak variation of cycle 21 PC2, and also quite strong variation of cycle 22 PC2. After that the situation changes such that Cycle 23 is still positive for Oulu as expected for the odd cycle, but slightly negative for Hermanus. Cycle 24 is again positive for both Oulu and Hermanus CR PC2 time series. There really is change from the patterns of earlier odd-even cycle pairs.  

In Figure \ref{fig:PC_timeseries_rest}a and b we show the PC1 and PC2 timeseries of the other station we have studied. They seem very similar to the PCs of Oulu CR for Cycles 20\,--\,24. Note, especially, that the positive phase (seen clearly in magnification) for Cycle 23 is slightly greater for Moscow and Thule than for the other CR stations. This is also the case for Thule in cycle 24, so it seems there is no bias just in cycle 23. Moscow had weird looking cosmic-ray data for the Cycle 24 , which would have distorted the whole PCA for Moscow. That is why we omitted the cycle 24 data for Moscow.

\begin{figure}
	\centering
	\includegraphics[width=1.0\textwidth]{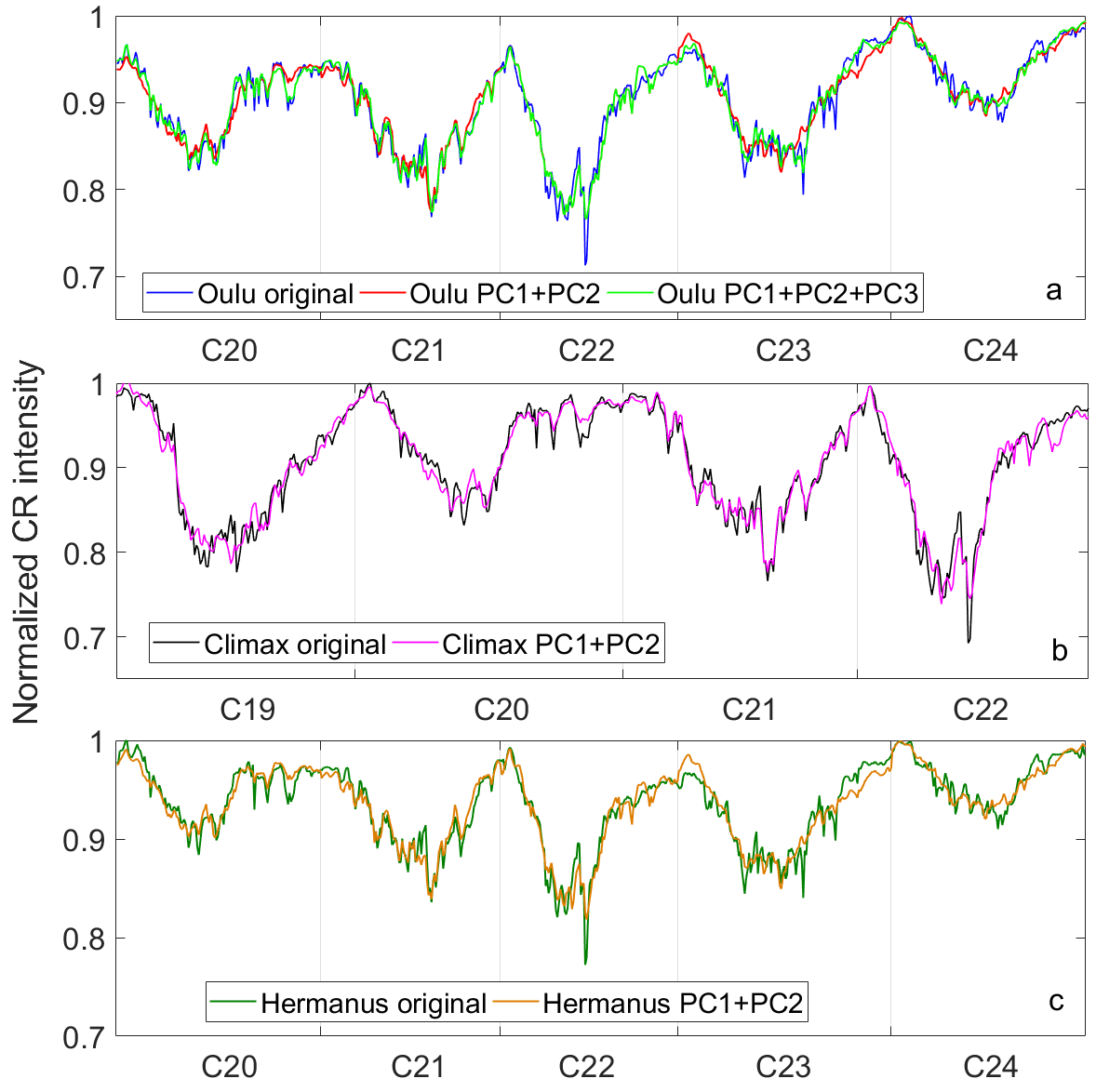}
		\caption{a) Original monthly CR time series for Oulu C20-C24 (blue), its PC1+PC2 proxy (red) and PC1+PC2+PC3 proxy (green) time series. b) Original monthly CR time series for Climax C19-C22 (black) and its PC1+PC2 proxy time series (magenta). c) Original monthly CR time series for Hermanus C20-C24 (green) and its PC1+PC2 proxy time series (brown).}
		\label{fig:CR_and_PC_proxy}
\end{figure}

Figure \ref{fig:CR_and_PC_proxy}a, b  and c show the original CR intensities and PC1+PC2 proxy time series for Oulu and Climax  and Hermanus stations, respectively. The main features are produced by these two main PCs quite well, especially for Climax data. The clearest differences are in the Cycle 23 for Oulu and Hermanus data. The late peaks and fast recovery after them are not brought forth by the PC1 and PC2. These features, which are unique for one cycle alone are produced by the higher order PCs. Cycle 23 has large value for EOF3 and when we add the PC3, i.e. plot also  PC1+PC2+PC3, the aforementioned features are corrected quite well (green curve in Fig. \ref{fig:CR_and_PC_proxy}a. Note that Cycle 22 does not change at all when PC3 is added and green PC1+PC2+PC3 proxy curve is on top the red line for C22. This is because its EOF3 for the C22 is practically zero. The case is similar to Hermanus station.

\begin{figure}
	\centering
	\includegraphics[width=1.0\textwidth]{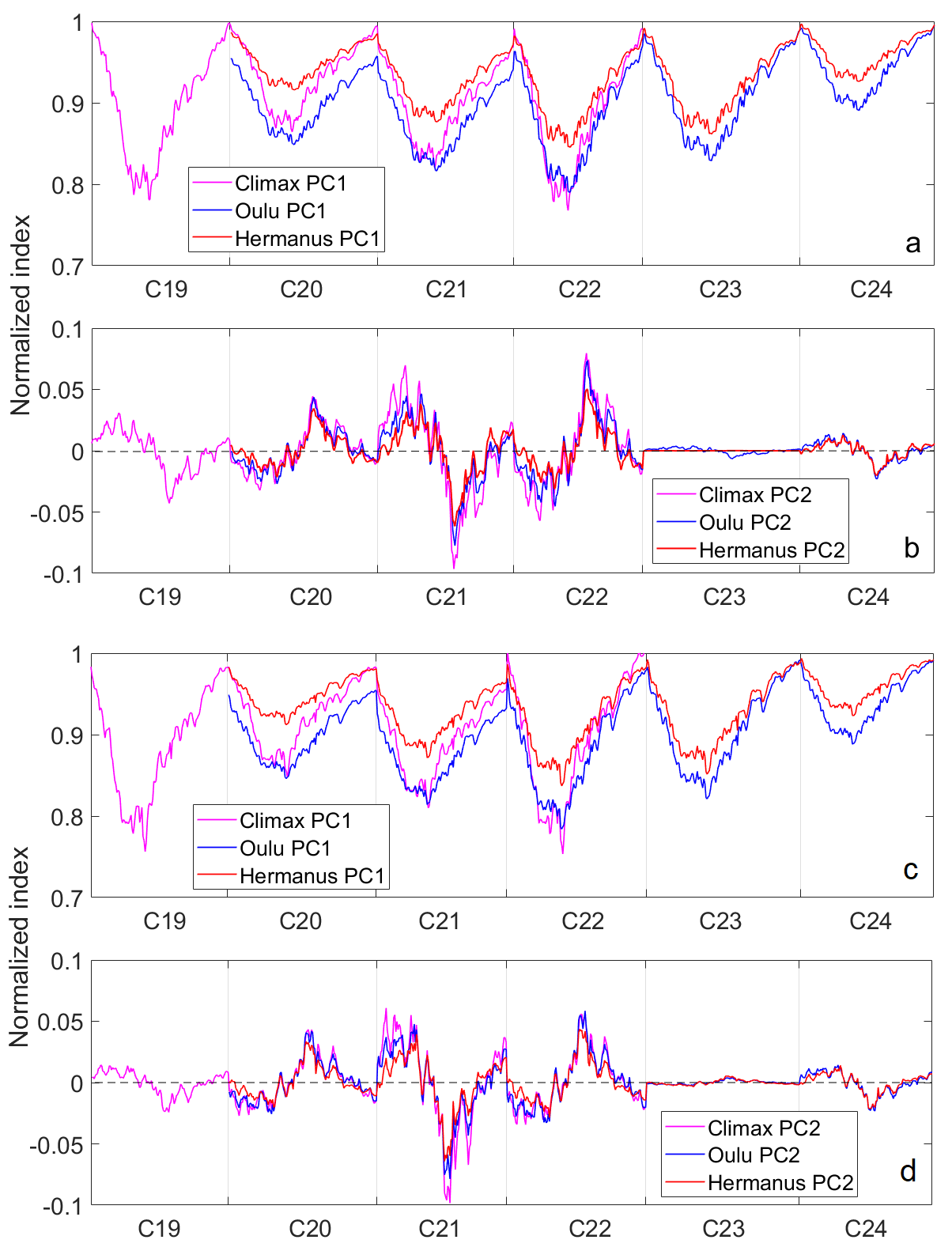}
		\caption{a) PC1 and b) PC2 time series of CR intensities for Oulu, Hermanus and Climax for Cycles 19\,--\,24 of the second PC analysis c) PC1 and d) PC2 time series of CR intensities for Cycles 19\,--\,24 of the third PC analysis.}
		\label{fig:PC_timeseries_second}
\end{figure}

In order to see if the choice of the starting minima (and the length of the cycle) of the cosmic-ray cycles affects to the PC analysis, we used the dates of the starting minima shown in Table 2. It is also a well-known fact that cosmic-ray flux lags the solar activity cycle from a few months to over a year depending on the station and cycle \citep{Mavromichalaki_1997, Gupta_2005, Rybansky_2009, Kane_2014, Porta_2018, Iskra_2019, Ross_2019, Koldobskiy_2022}. In the second analysis the PC1 and PC2 explain 78.1\,\% and 14.4\,\% of the total variation of the Oulu C20-C24 data, 74.9\% and 15.3\% of the variation of Hermanus C20-C24  and 74.5\,\% and 20.4\,\% of the Climax C19-C22 data. Hence the two main PCs account together almost the same part of the total variation of the CR data as in the fist analysis, i.e. 92.5\,\%, 90.2\% and 94.9\,\% for Oulu, Hermanus and Climax, respectively. Figures \ref{fig:PC_timeseries_second}a and b show the corresponding time series of PC1 and PC2, respectively. In the third analysis we used still larger time-lags and these starting dates are also available in the Table 2. In the third analysis the PC1 and PC2 explain 77.3\,\% and 15.2\,\% (together 92.5\,\%) of the total variation of the Oulu C20-C24 data, 73.3\% and 16.6\% (together (89,9\,\%) of the variation of Hermanus C20-C24  and 78.3\,\% and 17.5\,\% (together 95.8\,\%) of the Climax C19-C22 data. Figures \ref{fig:PC_timeseries_second}c and d show the corresponding time series of PC1 and PC2, respectively. Note the similarity of this figure compared to Figs. \ref{fig:PC_timeseries_first}a and b, and \ref{fig:PC_timeseries_second}a and b, although the cycle starting dates now differ from 4 to 14 months from the sunspot cycle starting dates. The most striking difference in the last PC2 compared to earlier PC2s is that even Oulu has now slightly negative phase for the odd Cycle 23. This result confirms the statement that Cycle 23 is more similar to earlier even cycles than odd cycles.

\begin{figure}
	\centering
	\includegraphics[width=1.0\textwidth]{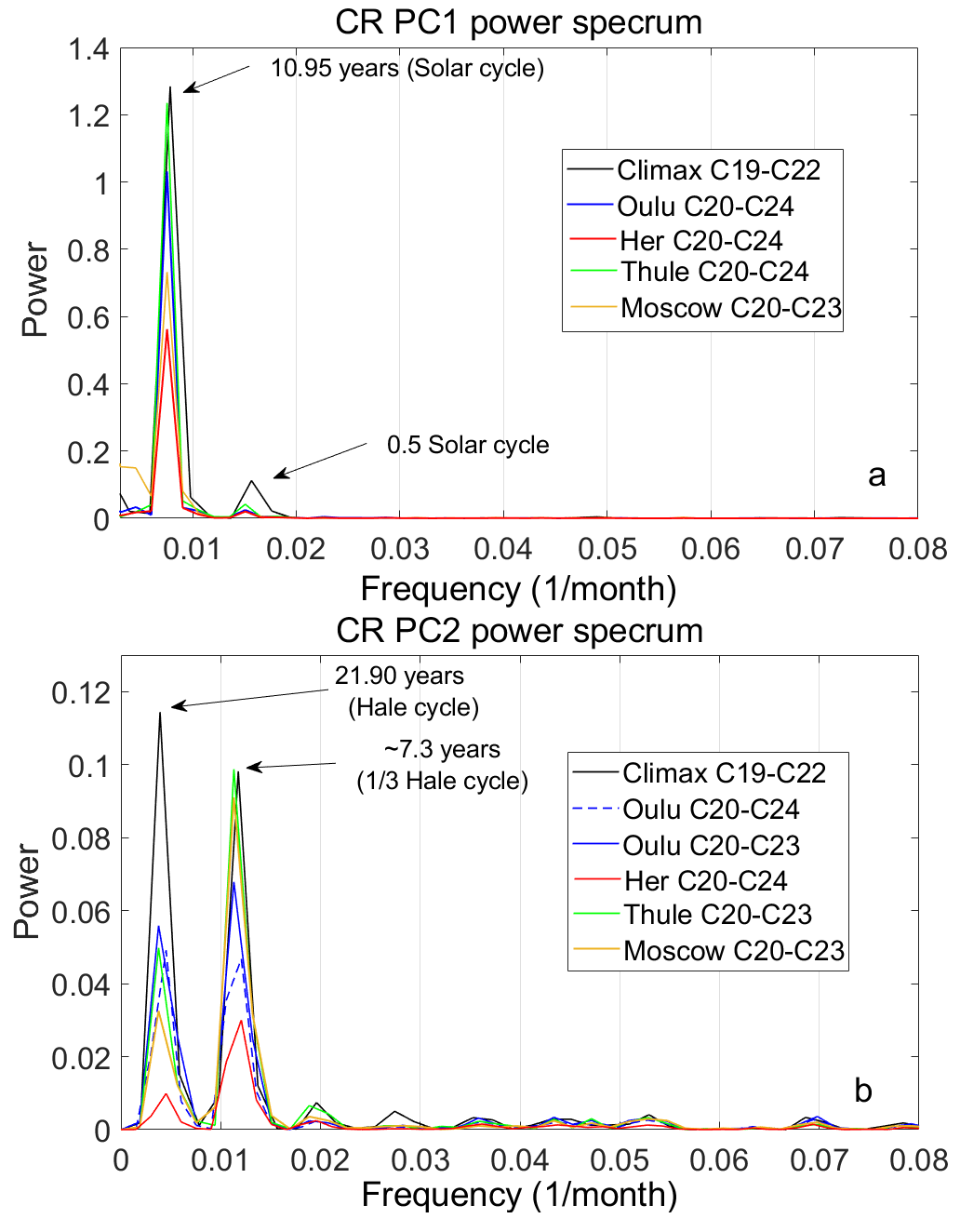}
		\caption{The power spectra of the a) PC1 and b) PC2 time series of Figure \ref{fig:PC_timeseries_first}.}
		\label{fig:PC1_PC2_power_spectra}
\end{figure}

Figure \ref{fig:PC1_PC2_power_spectra}a and b show the power spectra of PC1 and PC2 time series of Fig. \ref{fig:PC_timeseries_first}a, and b, respectively. It is clear that PCA effectively decomposes the solar cycle and Hale cycle related periods from the CR data. According to these spectra the solar cycle period is 10.95 years (on the average) and the Hale cycle about 21.9 years. In the Fig. \ref{fig:PC1_PC2_power_spectra}b, there is another smaller peak at period of one third of the Hale cycle. We believe that this peak is related to the phase change between the succeeding cycles. This causes the maxima (or minima) of the PC2 exist about 2/3 of the solar cycle (marked with two-headed arrows in Fig. \ref{fig:PC_timeseries_first}b). In the Fig. \ref{fig:PC1_PC2_power_spectra}b there are two spectra of Oulu PC2. The dashed blue spectral peaks are for the whole period C20\,--\,C24, and the solid blue spectral peaks for the case where the last Cycle 24, with "wrong phase" is omitted from the spectral analysis. We have added power spectra of Thule and Moscow for C20-C23 to the Fig. \ref{fig:PC1_PC2_power_spectra} to confirm that the peaks are the same (it is not necessary put all the power spectra, because Fig. \ref{fig:PC_timeseries_rest} shows that the PC1s and PC2 are very similar to all stations). The 1/3 of Hale cycle peak is even stronger than the whole cycle peak for Oulu, Hermanus, Thule and Moscow. We suppose that this is due to strong peaks for PC2 of the Cycles 20\,--\,22 for all these stations. However, this peak is somewhat artificial, because it comes from the concatenation of the separate PC2s of the different cycles. The important thing is that we have removed solar cycle related period from the CR data. Note that for Climax the whole Hale cycle peak is higher than the 1/3 Hale cycle peak, because Climax CR data consist of only the distinct Hale cycle pairs 19\,--\,20 and 21\,--\,22.
 
We can now compare our results with the earlier studies. The positive PC2, i.e positive phase in the first half of the cycle, means that the decrease in the CR intensity is slower than if it were negative. In Figs. \ref{fig:PC_timeseries_first} and \ref{fig:PC_timeseries_rest} we see that this is the case for odd cycles. This result is similar to those of, e.g., \cite{Mavromichalaki_1997} and \cite{VanAllen_2000}. For the the even Cycles 20 and 22 the first half of PC2 has negative phase and second half positive phase. This makes decreasing of CR cycle faster and recovering more rapid leading to more peak-like structure of the even cycles. That was the result of the earlier studies too (see \citep{Webber_1988, Mavromichalaki_1997, VanAllen_2000, Thomas_2014}). We have shown here the essential reason for these behaviors in CR, i.e. they are caused by the Hale cycle related component of the CR data.
\begin{figure}
	\centering
	\includegraphics[width=1.0\textwidth]{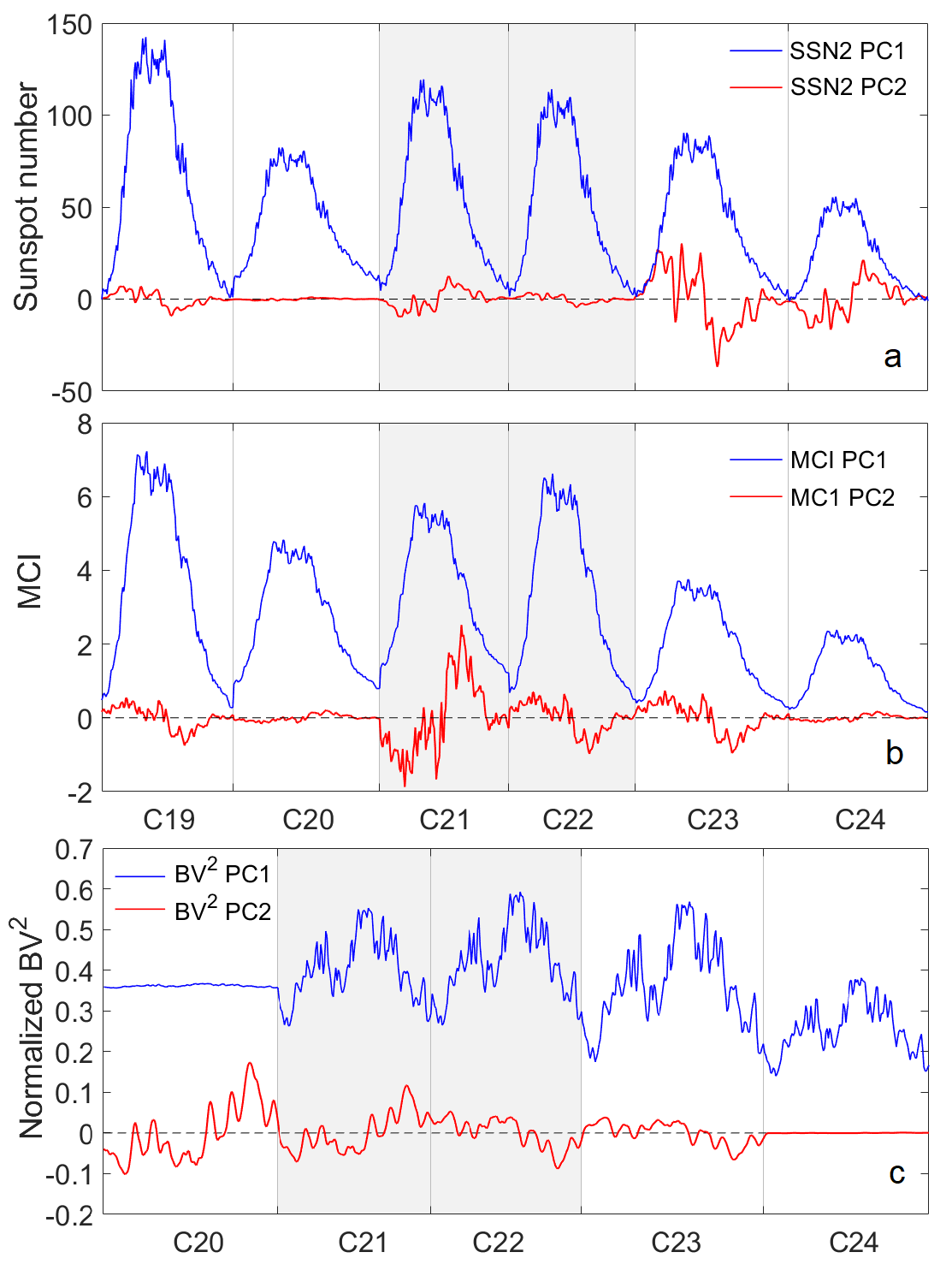}
		\caption{The PC1s and PC2s of a) sunspot numbers and b) Monthly corona indices for Solar Cycles 19\,--\,24. c) The PC1 and PC2 for Solar wind/IMF function Bv$^{2}$ of Solar Cycles 20-24.}
		\label{fig:SSN_CI_Bv2_proxy}
\end{figure}

We have also conducted the PC analysis for sunspot number 2.0 monthly data (SSN), solar monthly corona index (MCI), and function proportional to the IMF magnetic field intensity multiplied with square of solar-wind velocity Bv$^{2}$. The last one has been shown to  play the central role in the process of energy transfer \citep{Ahluwalia_2000, Verbanac_2011, Takalo_2021a}. Figure \ref{fig:SSN_CI_Bv2_proxy}a, b and c show the PC1 and PC2 timeseries for the SSN, MCI and Bv$^{2}$, respectively. Notice that OMNI data, which are used to calculate function Bv$^{2}$ exist only since 1964, i.e. for Solar Cycles 20-24. (The PC2s have been slightly smoothed to better see the phases of the cycles). Note that PC2s are in the same phase for all solar/IMF parameters. We would assume that the solar parameters anti-correlate with cosmic-ray data. Indeed, the PC2s of the cycles 21 and 22 (marked as gray) are exactly in opposite phase with the PC2 of the cosmic-ray data. Note that PC2 is by far strongest just for those cycles in CR data.
The solar wind speed and consequently function Bv$^{2}$ seem to be stronger always in the latter half of the cycle, i.e. during the descending phase of the Solar Cycle (see \ref{fig:SSN_CI_Bv2_proxy}c) \citep{Takalo_2021a}. We would then suppose that cosmic-ray has negative phase as preferred state in the second half of the cycle. This is actually the case for cycles 23 and 24 (instead slightly for Hermanus C23 and in one case for Oulu). Although we do not have solar wind data for the cycle 19, the situation could be the same for this cycle also. The cycle 20 is the only one which does not fit in to the aforementioned pattern. Note, however, that the Bv$^{2}$ PC1 is totally different for the Cycle 20 than the other cycles. The PC1 for Cycle 20 is almost constant and the stronger second half rather exists in the PC2 for this cycle. \cite{Agarwal_2008} already found that the interplanetary magnetic field strength (B) shows only a weak negative correlation (-0.35) with cosmic-rays for the solar cycle 20, whereas it shows a high anti-correlation for the solar cycles 21\,--\,23 (-0.76,-0.69). They suspect that a significant contribution to modulation from the termination shock during solar cycle 20 could dilute the correlation of cosmic-rays with the interplanetary magnetic field B at one astronomical unit for that cycle. The perturbations of the heliosphere are weaker and less widely spread during solar cycle 20 than during other solar cycles. This might lead to a situation where the heliospheric perturbations are relatively small for cosmic-ray particles allowing these particles to reach the Earth as if it was a minimum solar activity period. Note from figure \ref{fig:PC_timeseries_first}a that the decrease of the cosmic-ray intensity is smaller for Cycle 20 than for Cycle 23, especially for Hermanus, although sunspot numbers are almost the same for these cycles and Solar corona even stronger for Cycle 20 than for Cycle 23 (see Fig. \ref{fig:SSN_CI_Bv2_proxy}).

\subsection{Skewness and Kurtosis of the cycles}

\begin{figure}
	\centering
	\includegraphics[width=1.0\textwidth]{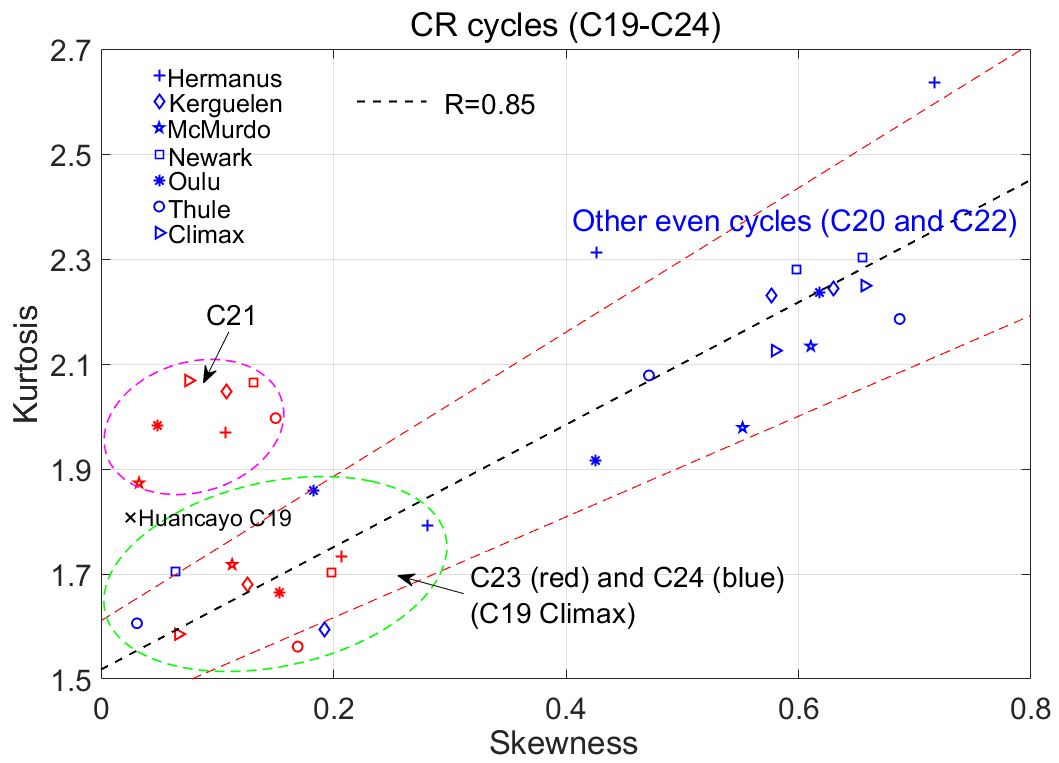}
		\caption{The skewness-kurtosis coordinate map showing these parameters for the Cycles 19-24 of all stations analysed in this study.}
		\label{fig:Skewness_kurtosis}
\end{figure}

We have calculated skewnesses and kurtoses of CR cycles 19-22 for Climax and cycles 20-24 for Hermanus, Kerguelen, McMurdo, Newark, Oulu and Thule. Figure \ref{fig:Skewness_kurtosis} shows the result in skewness-kurtosis (S\,--\,K) coordinate system. There are several interesting features. The cycles other than C21 are near the regression line (shown as black dashed line) such that cycles 23 and 24 are in the left lower corner meaning small skewness and low kurtosis. Interestingly Cycle 19 for Climax is among the cycles 23 and 24 in the left lower ellipse. The even cycles C20 and C22 are in the right upper corner with larger skewnesses and higher kurtoses. The correlation coefficient of the regression line is 0.85 and 95\% confidence limits are shown as red dashed lines. Note that only even Cycles 20 and 22 for Hermanus are somewhat outside the 95\% limits. The Cycles 21 are omitted from this regression and they are located compactly inside a small ellipse with small skewness and moderate kurtosis. We should remember that the Cycle 21 differed from other in PCA such that its weight for PC1 is smallest and PC2 highest of all cycles. Another classification could be that all odd cycles plus cycle 24 have small skewnesses and even cycles 20 and 22 have moderate to large skewnesses.
The skewness-kurtosis map also shows that Cycle 21 differs from others such that they are compactly located in a small area in this coordinate system. Also the skewness-kurtosis values for another odd Cycle 23 form a quite compact group, but with somewhat lower kurtoses and slightly larger skewnesses. The interesting thing is that the only Cycle 19 has almost smallest value for skewness and kurtosis of all cycles. It is a pity that we do not have any other results for the whole cycle 19, and even this data was measured with the old IGY-type monitor. There exists another whole Cycle 19 measured at Huancayo, and we have marked it as an 'x' to the skewness-kurtosis coordinate map. Its skewness is almost zero and it locates near the cycles 21 in the S\,--\,K map. The other cycles for Huancayo C20 and C21 are, however, so flat, i.e., have so heavy tails, that their kurtosis is outside the S\,--\,K map of Fig. \ref{fig:Skewness_kurtosis}. The reason is probably very high rigidity cutoff for that station, 13.4 Gev, and thus its insensitivity to low energy cosmic-rays \citep{Usoskin_2001, Usoskin_2017}. Another reason to omit Huancayo in PCA is that there exists only three whole cycles for this station.

\section{Conclusion}

We have studied CR fluxes of several stations using PC analyses. We present the PCAs of Climax CR for Cycles 19\,--\,23 and Hermanus and Oulu CR for Cycles 20\,--\,24 in more detail. We found that first and second PC explain 76.7\,\% and 18.8\,\%, 77.0\% and 13.2\% and 79.0\,\% and 13.3\,\% of the total variation of the Climax, Hermanus and Oulu CR flux data, respectively. This means that only the two main PCs account for 95.5\%, 90.2\% and 92.3\% of the total variation of Climax, Hermanus and Oulu CR intensity, respectively. The results for the time-lagged analyses (second and third PCA) give compatible results with similar significance of their PC1 and PC2. We have also shown that the PC1 is related to the solar cycle and PC2 is related to the Hale cycle, i.e. in the latter there is no solar cycle related peak in the power spectrum. According to PCA the normalized principal components are perpendicular towards each other. Note that almost one fifth of the Climax data are related to 22-year magnetic polarity cycle (for the second analysis over 20\,\%). This is likely due to the presence of grand solar maximum cycles in Climax data. For Hermanus and Oulu the percentage is lower, because there are the weak Cycles 23 and 24 included in the analysis. The polarity of the heliospheric magnetic field seems to be in very important role in the intensity of CR fluxes, at least during the grand solar maximum. We have also shown that especially PC2s of three solar related parameters SSN, SCI and Bv$^{2}$ are for the Cycles 21 and 22, i.e. for a whole odd-even Hale cycle, in the same phase with the cosmic-ray PC2s.

Our study shows that Cycle 23 is different from earlier odd cycles such that it has very weak positive phase in its PC2, and even slightly negative for southern hemisphere station Hermanus in the second and third analysis and for northern hemisphere Oulu in the third analysis. Interestingly, the Cycle 24 differs from the earlier even Cycles 20 and 22 such that it has positive phase in its PC2 for all analyses and this is even stronger than for odd Cycle 23. Furthermore, the Cycles 23 and 24 are very near each other in the skewness-kurtosis coordinate map with small skewness and low kurtosis values. On the other hand all other even cycles have larger skewnesses and higher kurtoses. The CR Cycles 21 differ from Cycles 23 such that their kurtoses are higher and they do not match the regression line, which combines the other odd and even cycles with correlation coefficient 0.85. The Cycle 19 is somewhat problematic with inadequate cosmic-ray data.

The aforementioned analysis leads us to conclude that the clear Hale cycle in the cosmic-ray data such that succeeding cycles are in different phase may be due only for solar cycle of the grand-maximum during the second half of 20$^{th}$ century \citep{Thomas_2014}. Another explanation could be that there is a phase shift, which started during the Cycle 23 \citep{McDonald_2010} and went on during Cycle 24 and the phases of even and odd cycles are changing places. We, however, rather suppose that while solar wind function Bv$^{2}$ has usually been stronger in the descending phase of the solar cycle (see Fig. \ref{fig:SSN_CI_Bv2_proxy}c PC1), due to high-speed streams \citep{Gosling_1977, Simon_1989, Cliver_1996, Echer_2004}, it is the cause for positive phase in the first half and negative phase in the second half of the PC2 for the weaker solar cycles like C23 and C24.

\begin{acknowledgments}
I acknowledge the NMDB database (www.nmdb.eu), \newline founded under the European Union’s FP7 programme (contract no. 213007), and the teams of the Climax, Hermanus, Kerguelen, McMurdo, Moscow, Newark, Oulu and Thule neutron monitors for providing data. The neutron-monitor data from McMurdo, Newark/Swarthmore, and Thule are provided by the University of Delaware Department of Physics and Astronomy and the Bartol Research Institute. Kerguelen neutron monitor data were kindly provided by Observatoire de Paris and the French Polar Institute (IPEV), France. Oulu NM is operated by Sodankyla Geophysical  Observatory of the University of Oulu. Moscow Neutron Monitor is operated by Pushkov Institute of Terrestrial  Magnetism, Ionosphere and radio wave propagation (IZMIRAN) of Russian Academy of Science. Hermanus NM is attended by South African National Space Agency. The SSN 2.0 data are fetched from www.bis.sidc.be/silso/datafiles, the monthly solar corona data from from www.ngdc.noaa.gov/stp/solar/corona.html, and the OMNI2 data from spdf.gsfc.nasa.gov/pub/data/omni/low\_res\_omni/. I am also grateful to prof. I. Usoskin for useful communication.
\end{acknowledgments}

\flushleft
\textbf{Disclosure of Potential Conflicts of Interest}
\footnotesize {The author declares that there are no conflicts of interest.}

\bibliographystyle{spr-mp-sola}
\bibliography{references_JT}  

\begin{thebibliography}{46}
\ifx\bisbn     \undefined \def\bisbn  #1{ISBN #1}\fi
\ifx\binits    \undefined \def\binits#1{#1}\fi
\ifx\bauthor   \undefined \def\bauthor#1{#1}\fi
\ifx\batitle   \undefined \def\batitle#1{#1}\fi
\ifx\bjtitle   \undefined \def\bjtitle#1{\textit{#1}}\fi
\ifx\bvolume   \undefined \def\bvolume#1{\textbf{#1}}\fi
\ifx\byear     \undefined \def\byear#1{#1}\fi
\ifx\bissue    \undefined \def\bissue#1{#1}\fi
\ifx\bfpage    \undefined \def\bfpage#1{#1}\fi
\ifx\blpage    \undefined \def\blpage #1{#1}\fi
\ifx\burl      \undefined \def\burl#1{#1}\fi
\ifx\href      \undefined \def\href#1#2{#2}\fi
\ifx\betal     \undefined \def\betal{et al.}\fi
\ifx\bctitle   \undefined \def\bctitle#1{#1}\fi
\ifx\beditor   \undefined \def\beditor#1{#1}\fi
\ifx\bbtitle   \undefined \def\bbtitle#1{\textit{#1}}\fi
\ifx\bedition  \undefined \def\bedition#1{#1}\fi
\ifx\bseriesno \undefined \def\bseriesno#1{\textbf{#1}}\fi
\ifx\blocation \undefined \def\blocation#1{#1}\fi
\ifx\bsertitle \undefined \def\bsertitle#1{\textit{#1}}\fi
\ifx\bsnm      \undefined \def\bsnm#1{#1}\fi
\ifx\bsuffix   \undefined \def\bsuffix#1{#1}\fi
\ifx\bparticle \undefined \def\bparticle#1{#1}\fi
\ifx\barticle  \undefined \def\barticle#1{}\fi
\ifx\binstitute  \undefined \def\binstitute#1{#1}\fi
\ifx\bpublisher  \undefined \def\bpublisher#1{#1}\fi
\ifx\doiurl    \undefined \def\doiurl#1{\href{#1}{DOI}}\fi
\makeatletter
\def\safeHref#1#2#3{\in@{http}{#2}\ifin@\href{#2}{#3}\else\href{#1#2}{#3}\fi}
\makeatother
\ifx\adsurl    \undefined
  \def\adsurl#1{\safeHref{https://ui.adsabs.harvard.edu/abs/}{#1}{ADS}}\fi
\ifx\arxivurl  \undefined
  \def\arxivurl#1{\safeHref{http://arxiv.org/abs/}{#1}{arXiv}}\fi
\ifx\botherref \undefined \def\botherref#1{}\fi
\ifx\url       \undefined \def\url#1{#1}\fi
\ifx\bchapter  \undefined \def\bchapter#1{}\fi
\ifx\bbook     \undefined \def\bbook#1{}\fi
\ifx\bcomment  \undefined \def\bcomment#1{#1}\fi
\ifx\oauthor   \undefined \def\oauthor#1{#1}\fi
\ifx\citeauthoryear \undefined\def \citeauthoryear#1{#1}\fi
\def\endbibitem {}
\ifx\bconflocation  \undefined \def\bconflocation#1{#1} \fi

\bibitem[\protect\citeauthoryear{{Agarwal} and {Mishra}}{2008}]{Agarwal_2008}
\begin{barticle}
\bauthor{\bsnm{{Agarwal}}, \binits{R.}},
\bauthor{\bsnm{{Mishra}}, \binits{R.K.}}:
\byear{2008},
\batitle{Solar cycle phenomena in cosmic ray intensity up to the recent solar
  cycle}.
\bjtitle{Physics Letters B}
\bvolume{664},
\bfpage{31}.
\doiurl{https://doi.org/10.1016/j.physletb.2008.04.057}.
\end{barticle}
\endbibitem

\bibitem[\protect\citeauthoryear{{Ahluwalia}}{2000}]{Ahluwalia_2000}
\begin{barticle}
\bauthor{\bsnm{{Ahluwalia}}, \binits{H.S.}}:
\byear{2000},
\batitle{{Galactic cosmic ray diurnal modulation, interplanetary magnetic field
  intensity and the planetary index Ap}}.
\bjtitle{Geophys.\ Res.\ Lett.}
\bvolume{27},
\bfpage{617}.
\doiurl{https://doi.org/10.1029/1999GL003716}.
\end{barticle}
\endbibitem

\bibitem[\protect\citeauthoryear{{Bhattacharyya} and
  {Okpala}}{2015}]{Bhattacharyya_2015}
\begin{barticle}
\bauthor{\bsnm{{Bhattacharyya}}, \binits{A.}},
\bauthor{\bsnm{{Okpala}}, \binits{K.C.}}:
\byear{2015},
\batitle{{Principal components of quiet time temporal variability of equatorial
  and low-latitude geomagnetic fields}}.
\bjtitle{J.\ Geophys.\ Res.}
\bvolume{120},
\bfpage{8799}.
\doiurl{https://doi.org/10.1002/2015JA021673}.
\end{barticle}
\endbibitem

\bibitem[\protect\citeauthoryear{{Bro} and {Smilde}}{2014}]{Bro_2014}
\begin{barticle}
\bauthor{\bsnm{{Bro}}, \binits{R.}},
\bauthor{\bsnm{{Smilde}}, \binits{A.K.}}:
\byear{2014},
\batitle{{Principal component analysis}}.
\bjtitle{Analytical Methods}
\bvolume{6},
\bfpage{2812}.
\end{barticle}
\endbibitem

\bibitem[\protect\citeauthoryear{{Broomhall}}{2017}]{Broomhall_2017}
\begin{barticle}
\bauthor{\bsnm{{Broomhall}}, \binits{A.N.}}:
\byear{2017},
\batitle{{A Helioseismic Perspective on the Depth of the Minimum Between Solar
  Cycles 23 and 24}}.
\bjtitle{Sol.\ Phys.}
\bvolume{292},
\bfpage{2812}.
\doiurl{https://doi.org/10.1007/s11207-017-1068-5}.
\end{barticle}
\endbibitem

\bibitem[\protect\citeauthoryear{{Cliver}, {Boriakoff}, and
  {Bounar}}{1996}]{Cliver_1996}
\begin{barticle}
\bauthor{\bsnm{{Cliver}}, \binits{E.W.}},
\bauthor{\bsnm{{Boriakoff}}, \binits{V.}},
\bauthor{\bsnm{{Bounar}}, \binits{K.H.}}:
\byear{1996},
\batitle{{The 22-year cycle of geomagnetic and solar wind activity}}.
\bjtitle{J.\ Geophys.\ Res.}
\bvolume{101},
\bfpage{27091}.
\doiurl{https://doi.org/10.1029/96JA02037}.
\adsurl{1996JGR...10127091C}.
\end{barticle}
\endbibitem

\bibitem[\protect\citeauthoryear{{Echer} et~al.}{2004}]{Echer_2004}
\begin{barticle}
\bauthor{\bsnm{{Echer}}, \binits{E.}},
\bauthor{\bsnm{{Gonzalez}}, \binits{W.D.}},
\bauthor{\bsnm{{Gonzalez}}, \binits{A.L.C.}},
\bauthor{\bsnm{{Prestes}}, \binits{A.}},
\bauthor{\bsnm{{Vieira}}, \binits{L.E.A.}},
\bauthor{\bsnm{{dal Lago}}, \binits{A.}},
\bauthor{\bsnm{{Guarnieri}}, \binits{F.L.}},
\bauthor{\bsnm{{Schuch}}, \binits{N.J.}}:
\byear{2004},
\batitle{{Long-term correlation between solar and geomagnetic activity}}.
\bjtitle{J. Atm.\ Sol.-Terr.\ Phys.}
\bvolume{66},
\bfpage{1019}.
\doiurl{https://doi.org/10.1016/j.jastp.2004.03.011}.
\adsurl{2004JASTP..66.1019E}.
\end{barticle}
\endbibitem

\bibitem[\protect\citeauthoryear{{Gnevyshev} and {Ohl}}{1948}]{Gnevyshev_1948}
\begin{barticle}
\bauthor{\bsnm{{Gnevyshev}}, \binits{M.N.}},
\bauthor{\bsnm{{Ohl}}, \binits{A.I.}}:
\byear{1948},
\batitle{{On the 22-year cycle of solar activity}}.
\bjtitle{Astron. Zh.}
\bvolume{25},
\bfpage{18}.
\end{barticle}
\endbibitem

\bibitem[\protect\citeauthoryear{{Gosling}, {Asbridge}, and
  {Bame}}{1977}]{Gosling_1977}
\begin{barticle}
\bauthor{\bsnm{{Gosling}}, \binits{J.T.}},
\bauthor{\bsnm{{Asbridge}}, \binits{J.R.}},
\bauthor{\bsnm{{Bame}}, \binits{S.J.}}:
\byear{1977},
\batitle{{An unusual aspect of solar wind speed variations during solar cycle
  20}}.
\bjtitle{J.\ Geophys.\ Res.}
\bvolume{82},
\bfpage{3311}.
\doiurl{https://doi.org/10.1029/JA082i022p03311}.
\adsurl{1977JGR....82.3311G}.
\end{barticle}
\endbibitem

\bibitem[\protect\citeauthoryear{{Gupta}, {Mishra}, and
  {Mishra}}{2005}]{Gupta_2005}
\begin{barticle}
\bauthor{\bsnm{{Gupta}}, \binits{M.}},
\bauthor{\bsnm{{Mishra}}, \binits{V.K.}},
\bauthor{\bsnm{{Mishra}}, \binits{A.P.}}:
\byear{2005},
\batitle{{Correlative study of solar activity and cosmic ray intensity for
  solar cycles 20 to 23}}.
\bjtitle{Proceedings of 29th International Cosmic Ray Conference Pune}
\bvolume{2},
\bfpage{147}.
\end{barticle}
\endbibitem

\bibitem[\protect\citeauthoryear{{Hannachi}, {Jolliffe}, and
  {Stephenson}}{2007}]{Hannachi_2007}
\begin{barticle}
\bauthor{\bsnm{{Hannachi}}, \binits{A.}},
\bauthor{\bsnm{{Jolliffe}}, \binits{I.T.}},
\bauthor{\bsnm{{Stephenson}}, \binits{D.B.}}:
\byear{2007},
\batitle{{Empirical orthogonal functions and related techniques in atmospheric
  science: A review}}.
\bjtitle{Int.\ J.\ Clim.}
\bvolume{27},
\bfpage{1119}.
\doiurl{https://doi.org/10.1002/joc.1499}.
\end{barticle}
\endbibitem

\bibitem[\protect\citeauthoryear{{Hempelmann} and
  {Weber}}{2012}]{Hempelmann_2012}
\begin{barticle}
\bauthor{\bsnm{{Hempelmann}}, \binits{A.}},
\bauthor{\bsnm{{Weber}}, \binits{W.}}:
\byear{2012},
\batitle{{Correlation Between the Sunspot Number, the Total Solar Irradiance,
  and the Terrestrial Insolation}}.
\bjtitle{Sol.\ Phys.}
\bvolume{277},
\bfpage{417}.
\doiurl{https://doi.org/10.1007/s11207-011-9905-4}.
\end{barticle}
\endbibitem

\bibitem[\protect\citeauthoryear{{Holappa}, {Mursula}, and
  {Asikainen}}{2014}]{Holappa_2014_2}
\begin{barticle}
\bauthor{\bsnm{{Holappa}}, \binits{L.}},
\bauthor{\bsnm{{Mursula}}, \binits{K.}},
\bauthor{\bsnm{{Asikainen}}, \binits{T.}}:
\byear{2014},
\batitle{{A new method to estimate annual solar wind parameters and
  contributions of different solar wind structures to geomagnetic activity}}.
\bjtitle{J.\ Geophys.\ Res.}
\bvolume{119},
\bfpage{9407}.
\doiurl{https://doi.org/10.1002/2014JA020599}.
\adsurl{2014JGRA..119.9407H}.
\end{barticle}
\endbibitem

\bibitem[\protect\citeauthoryear{{Holappa} et~al.}{2014}]{Holappa_2014_1}
\begin{barticle}
\bauthor{\bsnm{{Holappa}}, \binits{L.}},
\bauthor{\bsnm{{Mursula}}, \binits{K.}},
\bauthor{\bsnm{{Asikainen}}, \binits{T.}},
\bauthor{\bsnm{{Richardson}}, \binits{I.G.}}:
\byear{2014},
\batitle{{Annual fractions of high-speed streams from principal component
  analysis of local geomagnetic activity}}.
\bjtitle{J.\ Geophys.\ Res.}
\bvolume{119},
\bfpage{4544}.
\doiurl{https://doi.org/10.1002/2014JA019958}.
\adsurl{2014JGRA..119.4544H}.
\end{barticle}
\endbibitem

\bibitem[\protect\citeauthoryear{{Iskra} et~al.}{2019}]{Iskra_2019}
\begin{botherref}
\oauthor{\bsnm{{Iskra}}, \binits{K.}},
\oauthor{\bsnm{{Siluszyk}}, \binits{M.}},
\oauthor{\bsnm{{Alania}}, \binits{M.}},
\oauthor{\bsnm{{Wozniak}}, \binits{W.}}:
2019,
{Experimental Investigation of the Delay Time in Galactic Cosmic Ray Flux in
  Different Epochs of Solar Magnetic Cycles: 1959 – 2014}.
\textit{Sol.\ Phys.}
\textbf{294}.
\end{botherref}
\endbibitem

\bibitem[\protect\citeauthoryear{{Kane}}{2014}]{Kane_2014}
\begin{barticle}
\bauthor{\bsnm{{Kane}}, \binits{R.P.}}:
\byear{2014},
\batitle{{Lags and Hysteresis Loops of Cosmic Ray Intensit yVersus Sunspot
  Numbers: Quantitative Estimates for Cycles 19 – 23 and a Preliminary
  Indicationfor Cycle 24}}.
\bjtitle{Sol.\ Phys.}
\bvolume{289},
\bfpage{2727}.
\doiurl{https://doi.org/10.1007/s11207-014-0479-9}.
\end{barticle}
\endbibitem

\bibitem[\protect\citeauthoryear{{Koldobskiy} et~al.}{2022}]{Koldobskiy_2022}
\begin{barticle}
\bauthor{\bsnm{{Koldobskiy}}, \binits{S.A.}},
\bauthor{\bsnm{{K{\"a}hk{\"o}nen}}, \binits{R.}},
\bauthor{\bsnm{{Hofer}}, \binits{B.}},
\bauthor{\bsnm{{Krivova}}, \binits{N.A.}},
\bauthor{\bsnm{{Kovaltsov}}, \binits{G.A.}},
\bauthor{\bsnm{{Usoskin}}, \binits{I.G.}}:
\byear{2022},
\batitle{{Time Lag Between Cosmic-Ray and Solar Variability: Sunspot Numbers
  and Open Solar Magnetic Flux}}.
\bjtitle{Sol.\ Phys.}
\bvolume{297}.
\doiurl{https://doi.org/10.1007/s11207-022-01970-1}.
\end{barticle}
\endbibitem

\bibitem[\protect\citeauthoryear{{Krishnamoorthy}}{2006}]{Krishnamoorthy_2006}
\begin{bbook}
\bauthor{\bsnm{{Krishnamoorthy}}, \binits{K.}}:
\byear{2006},
\bbtitle{Handbook of statistical distributions with applications},
\bpublisher{Chapman \& Hall/CRC, Taylor \& Francis Group, Boca Raton, FL
  33487-2742}.
\end{bbook}
\endbibitem

\bibitem[\protect\citeauthoryear{{Kumar}, {Rai}, and
  {Kumar}}{2008}]{Kumar_2008}
\begin{barticle}
\bauthor{\bsnm{{Kumar}}, \binits{D.}},
\bauthor{\bsnm{{Rai}}, \binits{C.S.}},
\bauthor{\bsnm{{Kumar}}, \binits{S.}}:
\byear{2008},
\batitle{{Principal Component Analysis for Data Compression and Face
  Recognition}}.
\bjtitle{INFOCOMP Journal of Computer Science}
\bvolume{7},
\bfpage{48}.
\end{barticle}
\endbibitem

\bibitem[\protect\citeauthoryear{{Lin}}{2012}]{Lin_2012}
\begin{barticle}
\bauthor{\bsnm{{Lin}}, \binits{J.-W.}}:
\byear{2012},
\batitle{{Ionospheric total electron content seismo-perturbation after Japan's
  March 11, 2011, M=9.0 Tohoku earthquake under a geomagnetic storm; a
  nonlinear principal component analysis}}.
\bjtitle{Astrophys. Space Sci.}
\bvolume{341},
\bfpage{251}.
\doiurl{https://doi.org/10.1007/s10509-012-1128-0}.
\end{barticle}
\endbibitem

\bibitem[\protect\citeauthoryear{{Makarov}}{1994}]{Makarov_1994}
\begin{barticle}
\bauthor{\bsnm{{Makarov}}, \binits{V.I.}}:
\byear{1994},
\batitle{{Global magnetic activity in 22-year solar cycles}}.
\bjtitle{Sol.\ Phys.}
\bvolume{150},
\bfpage{359}.
\end{barticle}
\endbibitem

\bibitem[\protect\citeauthoryear{{Mavromichalaki}
  et~al.}{1997}]{Mavromichalaki_1997}
\begin{barticle}
\bauthor{\bsnm{{Mavromichalaki}}, \binits{H.}},
\bauthor{\bsnm{{Belehaki}}, \binits{A.}},
\bauthor{\bsnm{{Rafios}}, \binits{X.}},
\bauthor{\bsnm{{Tsagouri}}, \binits{I.}}:
\byear{1997},
\batitle{{Hale-cycle effects in cosmic-ray intensity during the last four
  cycles}}.
\bjtitle{Astrophys. Space Sci.}
\bvolume{246},
\bfpage{7}.
\end{barticle}
\endbibitem

\bibitem[\protect\citeauthoryear{{McDonald}, {Webber}, and
  {Reames}}{2010}]{McDonald_2010}
\begin{botherref}
\oauthor{\bsnm{{McDonald}}, \binits{F.B.}},
\oauthor{\bsnm{{Webber}}, \binits{W.}},
\oauthor{\bsnm{{Reames}}, \binits{D.V.}}:
2010,
{Unusual time histories of galactic and anomalous cosmic rays at 1 AU over the
  deep solar minimum of cycle 23/24}.
\textit{Geophys.\ Res.\ Lett.}
\textbf{37}.
\end{botherref}
\endbibitem

\bibitem[\protect\citeauthoryear{{Mishra}, R., and
  {Tiwari}}{2008}]{Mishra_2008}
\begin{botherref}
\oauthor{\bsnm{{Mishra}}, \binits{R.K.}},
\oauthor{\bsnm{R.}, \binits{A.}},
\oauthor{\bsnm{{Tiwari}}, \binits{S.}}:
2008,
{Solar cycle variation of cosmic ray intensity along with interplanetary and
  solar wind plasma parameters}.
\textit{Latvian J. of Phys. and Techn. Sci}.
\doiurl{https://doi.org/0.2478/v10047-008-0013-7}.
\end{botherref}
\endbibitem

\bibitem[\protect\citeauthoryear{{Okpala} and {Okeke}}{2014}]{Okpala_2014}
\begin{bchapter}
\bauthor{\bsnm{{Okpala}}, \binits{K.}},
\bauthor{\bsnm{{Okeke}}, \binits{F.}}:
\byear{2014},
\bctitle{{Variability of the Daily Cosmic Ray Count rates in the Northern
  Hemisphere}}.
In: \bbtitle{40th COSPAR Scientific Assembly}
\bseriesno{40},
\bfpage{D1.3}.
\end{bchapter}
\endbibitem

\bibitem[\protect\citeauthoryear{{Oloketuyi} et~al.}{2020}]{Oloketuyi_2020}
\begin{barticle}
\bauthor{\bsnm{{Oloketuyi}}, \binits{J.}},
\bauthor{\bsnm{{Liu}}, \binits{Y.}},
\bauthor{\bsnm{{Amanambu}}, \binits{A.C.}},
\bauthor{\bsnm{{Zhao}}, \binits{M.}}:
\byear{2020},
\batitle{{Responses and Periodic Variations of Cosmic Ray Intensity and Solar
  Wind Speed to Sunspot Numbers}}.
\bjtitle{Advances in Astronomy}
\bvolume{2020}.
\doiurl{https://doi.org/10.1155/2020/3527570}.
\adsurl{1995SoPh..156..179O}.
\end{barticle}
\endbibitem

\bibitem[\protect\citeauthoryear{{Owens} et~al.}{2015}]{Owens_2015}
\begin{barticle}
\bauthor{\bsnm{{Owens}}, \binits{M.J.}},
\bauthor{\bsnm{{McCracken}}, \binits{K.G.}},
\bauthor{\bsnm{{Lockwood}}, \binits{M.}},
\bauthor{\bsnm{{Barnard}}, \binits{L.}}:
\byear{2015},
\batitle{{The heliospheric Hale cycle over the last 300 years and its
  implications for a "lost" late 18th century solar cycle}}.
\bjtitle{J.\ Space\ Weather\ Space\ Clim.}
\bvolume{5},
\bfpage{A30}.
\doiurl{https://doi.org/10.1051/swsc/2015032}.
\adsurl{2015JSWSC...5A..30O}.
\end{barticle}
\endbibitem

\bibitem[\protect\citeauthoryear{{Porta}}{2018}]{Porta_2018}
\begin{barticle}
\bauthor{\bsnm{{Porta}}, \binits{R.D.}}:
\byear{2018},
\batitle{Wavelet-based transformations for nonlinear signal processing}.
\bjtitle{Astrophys. Space Sci.}
\bvolume{363}.
\doiurl{https://doi.org/10.1007/s10509-018-3360-8}.
\end{barticle}
\endbibitem

\bibitem[\protect\citeauthoryear{{Ross} and {Chaplin}}{2019}]{Ross_2019}
\begin{barticle}
\bauthor{\bsnm{{Ross}}, \binits{E.}},
\bauthor{\bsnm{{Chaplin}}, \binits{W.J.}}:
\byear{2019},
\batitle{{The Behaviour of Galactic Cosmic-Ray Intensity During Solar Activity
  Cycle 24}}.
\bjtitle{Sol.\ Phys.}
\bvolume{294},
\bfpage{18203}.
\doiurl{https://doi.org//10.1007/s11207-019-1397-7}.
\end{barticle}
\endbibitem

\bibitem[\protect\citeauthoryear{{Rybansky}, {Kudela}, and
  {Minarovjech}}{2009}]{Rybansky_2009}
\begin{botherref}
\oauthor{\bsnm{{Rybansky}}, \binits{M.}},
\oauthor{\bsnm{{Kudela}}, \binits{K.}},
\oauthor{\bsnm{{Minarovjech}}, \binits{M.}}:
2009,
{Solar corona and cosmic rays 1953 - 2008}.
\textit{Proceedings of the 31st ICRC}.
\end{botherref}
\endbibitem

\bibitem[\protect\citeauthoryear{{Savić} et~al.}{2019}]{Savic_2019}
\begin{botherref}
\oauthor{\bsnm{{Savić}}, \binits{M.}},
\oauthor{\bsnm{{Dragić}}, \binits{A.}},
\oauthor{\bsnm{D.}, \binits{M.}},
\oauthor{\bsnm{N.}, \binits{V.}},
\oauthor{\bsnm{{Banjanac}}, \binits{R.}},
\oauthor{\bsnm{{Joković}}, \binits{D.}},
\oauthor{\bsnm{{Udovičić}}, \binits{V.}}:
2019,
{A novel method for atmospheric correction of cosmic-ray data based on
  principal component analysis}.
\textit{Astroparticle Physics},
1.
\end{botherref}
\endbibitem

\bibitem[\protect\citeauthoryear{{Simon} and {Legrand}}{1989}]{Simon_1989}
\begin{barticle}
\bauthor{\bsnm{{Simon}}, \binits{P.A.}},
\bauthor{\bsnm{{Legrand}}, \binits{J.P.}}:
\byear{1989},
\batitle{{Solar cycle and geomagnetic activity: a review for geophysicists.
  Part 2. The solar sources of geomagnetic activity and their links with
  sunspot cycle activity.}}
\bjtitle{Ann.\ Geophys.}
\bvolume{7},
\bfpage{579}.
\adsurl{1989AnGeo...7..579S}.
\end{barticle}
\endbibitem

\bibitem[\protect\citeauthoryear{{Takalo}}{2021a}]{Takalo_2021a}
\begin{barticle}
\bauthor{\bsnm{{Takalo}}, \binits{J.}}:
\byear{2021}a,
\batitle{{Comparison of geomagnetic indices during even and odd solar cycles
  SC17-SC24: Signatures of Gnevyshev gap in geomagnetic activity}}.
\bjtitle{Sol.\ Phys.}
\bvolume{296}.
\doiurl{https://doi.org/10.1007/s11207-021-01765-w}.
\end{barticle}
\endbibitem

\bibitem[\protect\citeauthoryear{{Takalo}}{2021b}]{Takalo_2021b}
\begin{barticle}
\bauthor{\bsnm{{Takalo}}, \binits{J.}}:
\byear{2021}b,
\batitle{{Separating the aa-index into Solar and Hale Cycle Related Components
  Using Principal Component Analysis}}.
\bjtitle{Sol.\ Phys.}
\bvolume{296}.
\doiurl{https://doi.org/10.1007/s11207-021-01825-1}.
\end{barticle}
\endbibitem

\bibitem[\protect\citeauthoryear{{Takalo} and {Mursula}}{2018}]{Takalo_2018}
\begin{barticle}
\bauthor{\bsnm{{Takalo}}, \binits{J.}},
\bauthor{\bsnm{{Mursula}}, \binits{K.}}:
\byear{2018},
\batitle{{Principal component analysis of sunspot cycle shape}}.
\bjtitle{Astron. Astrophys.}
\bvolume{620 A100}.
\doiurl{https://doi.org/10.1051/0004-6361/201833924}.
\end{barticle}
\endbibitem

\bibitem[\protect\citeauthoryear{{Takalo} and {Mursula}}{2020}]{Takalo_2020a}
\begin{barticle}
\bauthor{\bsnm{{Takalo}}, \binits{J.}},
\bauthor{\bsnm{{Mursula}}, \binits{K.}}:
\byear{2020},
\batitle{{Comparison of the shape and temporal evolution of even and odd solar
  cycles}}.
\bjtitle{Astron. Astrophys.}
\bvolume{636 A11}.
\doiurl{https://doi.org/10.1051/0004-6361/202037488}.
\end{barticle}
\endbibitem

\bibitem[\protect\citeauthoryear{{Thomas}, {Owens}, and {Lockwood,
  M.}}{2014}]{Thomas_2014}
\begin{barticle}
\bauthor{\bsnm{{Thomas}}, \binits{S.R.}},
\bauthor{\bsnm{{Owens}}, \binits{M.J.}},
\bauthor{\bsnm{{Lockwood, M.}}}:
\byear{2014},
\batitle{{The 22-Year Hale Cycle in Cosmic Ray Flux – Evidence for Direct
  Heliospheric Modulation}}.
\bjtitle{Sol.\ Phys.}
\bvolume{289},
\bfpage{407}.
\doiurl{https://doi.org/10.1007/s11207-013-0341-5}.
\end{barticle}
\endbibitem

\bibitem[\protect\citeauthoryear{{Thomas} et~al.}{2017}]{Thomas_2017}
\begin{barticle}
\bauthor{\bsnm{{Thomas}}, \binits{S.}},
\bauthor{\bsnm{{Owens}}, \binits{M.}},
\bauthor{\bsnm{{Lockwood}}, \binits{M.}},
\bauthor{\bsnm{{Owen}}, \binits{C.}}:
\byear{2017},
\batitle{{Decadal trends in the diurnal variation of galactic cosmic rays
  observed using neutron monitor data}}.
\bjtitle{Ann.\ Geophys.}
\bvolume{35},
\bfpage{825}.
\doiurl{https://doi.org/10.5194/angeo-35-825-2017}.
\adsurl{2017AnGeo..35..825T}.
\end{barticle}
\endbibitem

\bibitem[\protect\citeauthoryear{{Usoskin} et~al.}{2001}]{Usoskin_2001}
\begin{barticle}
\bauthor{\bsnm{{Usoskin}}, \binits{I.G.}},
\bauthor{\bsnm{{Bobik}}, \binits{P.}},
\bauthor{\bsnm{{Gladysheva}}, \binits{O.G.}},
\bauthor{\bsnm{{Kananen}}, \binits{H.}},
\bauthor{\bsnm{{Kovaltsov}}, \binits{G.A.}},
\bauthor{\bsnm{{Kudela}}, \binits{K.}}:
\byear{2001},
\batitle{{Sensitivity of a neutron monitor to galactic cosmic rays}}.
\bjtitle{Adv.\ Space\ Res.}
\bvolume{27},
\bfpage{565}.
\doiurl{https://doi.org/10.1016/S0273-1177(01)00094-1}.
\adsurl{2001AdSpR..27..565U}.
\end{barticle}
\endbibitem

\bibitem[\protect\citeauthoryear{{Usoskin} et~al.}{2017}]{Usoskin_2017}
\begin{barticle}
\bauthor{\bsnm{{Usoskin}}, \binits{I.G.}},
\bauthor{\bsnm{{Gil}}, \binits{A.}},
\bauthor{\bsnm{{Kovaltsov}}, \binits{G.A.}},
\bauthor{\bsnm{{Mishev}}, \binits{A.L.}},
\bauthor{\bsnm{{Mikhailov}}, \binits{V.V.}}:
\byear{2017},
\batitle{{Heliospheric modulation of cosmic rays during the neutron monitor
  era: Calibration using PAMELA data for 2006--2010}}.
\bjtitle{J.\ Geophys.\ Res.}
\bvolume{122},
\bfpage{3875}.
\doiurl{https://doi.org/10.1002/2016JA023819}.
\end{barticle}
\endbibitem

\bibitem[\protect\citeauthoryear{{V{\"a}is{\"a}nen}, {Usoskin}, and
  {Mursula}}{2021}]{Vaisanen_2021}
\begin{barticle}
\bauthor{\bsnm{{V{\"a}is{\"a}nen}}, \binits{P.}},
\bauthor{\bsnm{{Usoskin}}, \binits{I.}},
\bauthor{\bsnm{{Mursula}}, \binits{K.}}:
\byear{2021},
\batitle{Seven decades of neutron monitors (1951–2019): Overview and
  evaluation of data sources}.
\bjtitle{J.\ Geophys.\ Res.}
\bvolume{126},
\bfpage{e2020JA028941}.
\doiurl{https://doi.org/10.1029/2020JA028941}.
\end{barticle}
\endbibitem

\bibitem[\protect\citeauthoryear{{Van Allen}}{2000}]{VanAllen_2000}
\begin{barticle}
\bauthor{\bsnm{{Van Allen}}, \binits{J.A..}}:
\byear{2000},
\batitle{{On the Modulation of Galactic Cosmic Ray Intensity during Solar
  Activity Cycles 19, 20, 21, 22 and Early 23}}.
\bjtitle{Geophys.\ Res.\ Lett.}
\bvolume{27},
\bfpage{2453}.
\end{barticle}
\endbibitem

\bibitem[\protect\citeauthoryear{{Verbanac} et~al.}{2011}]{Verbanac_2011}
\begin{barticle}
\bauthor{\bsnm{{Verbanac}}, \binits{G.}},
\bauthor{\bsnm{{Vr{\v{s}}nak}}, \binits{B.}},
\bauthor{\bsnm{{{\v{Z}}ivkovi{\'c}}}, \binits{S.}},
\bauthor{\bsnm{{Hojsak}}, \binits{T.}},
\bauthor{\bsnm{{Veronig}}, \binits{A.M.}},
\bauthor{\bsnm{{Temmer}}, \binits{M.}}:
\byear{2011},
\batitle{{Solar wind high-speed streams and related geomagnetic activity in the
  declining phase of solar cycle 23}}.
\bjtitle{Astron. Astrophys.}
\bvolume{533},
\bfpage{A49}.
\doiurl{https://doi.org/10.1051/0004-6361/201116615}.
\end{barticle}
\endbibitem

\bibitem[\protect\citeauthoryear{{Webber} and {Lockwood}}{1988}]{Webber_1988}
\begin{barticle}
\bauthor{\bsnm{{Webber}}, \binits{W.R.}},
\bauthor{\bsnm{{Lockwood}}, \binits{J.A.}}:
\byear{1988},
\batitle{{Characteristics of the 22-year modulation of cosmic rays as seen by
  neutron monitors}}.
\bjtitle{J.\ Geophys.\ Res.}
\bvolume{93},
\bfpage{8735}.
\end{barticle}
\endbibitem

\bibitem[\protect\citeauthoryear{{Wilson}}{1988}]{Wilson_1988}
\begin{barticle}
\bauthor{\bsnm{{Wilson}}, \binits{R.M.}}:
\byear{1988},
\batitle{{Bimodality and the Hale cycle}}.
\bjtitle{Sol.\ Phys.}
\bvolume{117},
\bfpage{269–}.
\end{barticle}
\endbibitem

\bibitem[\protect\citeauthoryear{{Zharkova} et~al.}{2015}]{Zharkova_2015}
\begin{barticle}
\bauthor{\bsnm{{Zharkova}}, \binits{V.V.}},
\bauthor{\bsnm{{Shepherd}}, \binits{S.J.}},
\bauthor{\bsnm{{Popova}}, \binits{E.}},
\bauthor{\bsnm{{Zharkov}}, \binits{S.I.}}:
\byear{2015},
\batitle{{Heartbeat of the Sun from Principal Component Analysis and prediction
  of solar activity on a millenium timescale}}.
\bjtitle{Nature\ Sci.\ Rep.}
\bvolume{5},
\bfpage{15689}.
\doiurl{https://doi.org/10.1038/srep15689}.
\adsurl{2015NatSR...515689Z}.
\end{barticle}
\endbibitem

\end{thebibliography}


\end{document}